\documentstyle[12pt,epsfig]{article}
 1
 1
 1

\begin{document}

\newcommand{\D}{\mbox{D}}
\newcommand{\barF}{\bar{F}}
\newcommand{\id}{\mbox{1$\!\!$I}}
\newcommand{\be}{\begin{equation}}
\newcommand{\ee}{\end{equation}}
\newcommand{\bea}{\begin{eqnarray}}
\newcommand{\eea}{\end{eqnarray}}
\newcommand{\tp}{\tilde \pi}
\newcommand{\ts}{\tilde \sigma}
\newcommand{\tchi}{\tilde \chi}
\newcommand{\nn}{\nonumber}
\newcommand{\no}{\nonumber\\}
\newcommand{\adot}{\stackrel{.}{\alpha}}
\newcommand{\bdot}{\stackrel{.}{\beta}}
\newcommand{\dal}{ \frac{ \partial }{ \partial \theta^\alpha } }
\newcommand{\dbe}{ \frac{ \partial }{ \partial \theta^\beta } }
\newcommand{\dadot}{ \frac{ \partial }{ \partial \bar \theta_{\adot} } }
\newcommand{\dbdot}{ \frac{ \partial }{ \partial \bar \theta_{\bdot} } }
\newcommand{\dedouble}{ \stackrel{ \leftrightarrow }{ \partial } }
    
\newcommand{\km}{{\bf k}}  
\newcommand{\kmp}{{\bf k}^{\prime}}  
\newcommand{\ak}{a_{\km}}  
\newcommand{\bk}{b_{\km}}  
\newcommand{\akp}{a_{\km^{\prime}}}  
\newcommand{\bkp}{b_{\km^{\prime}}}  

\renewcommand{\Box}{box}
\newcommand{\unityT}{{\bf \id}_{\bf T}}
\newcommand{\unityG}{\id_{\Gamma}}
\newcommand{\ptensor}{\otimes}
\newcommand{\bareta}{\bar{\eta}}
\newcommand{\barlambda}{\bar{\lambda}}
\newcommand{\barT}{\overline{T}}
\newcommand{\dslash}{\gamma_{\mu}\partial^{\mu}}
\newcommand{\pslash}{\gamma_{\mu} p^{\mu}}
\newcommand{\munu}{(\mu \leftrightarrow \nu)}
\newcommand{\albe}{(\alpha \leftrightarrow \beta)}
\begin{titlepage}

\noindent
\begin{center}
{\Large \bf Model independent constraints on leptoquarks\\
\vspace{0.5cm}
from $\mu$ and $\tau$ lepton rare processes}
\end{center}
\bigskip 
\begin{center}
{\large Emidio Gabrielli}
\end{center}

\begin{center}
{\it Departamento de F\'\i sica Te\'orica, C-XI \\ 
Instituto de F\'\i sica Te\'orica, C-XVI\\
Universidad Aut\'onoma, E-28049 Madrid, Spain}
\end{center}

\bigskip
\begin{center}
{\bf Abstract}\\
\end{center}

We perform a model independent analysis so as to constrain
the leptoquark (LQ) models from negative searches for
$\mu \to e \gamma$, $\mu \to 3e$ decays
(and analogous processes in the $\tau$ sector), 
and coherent $\mu-e$ conversion in nuclei.
We considerably improve some constraints
obtained by analyses known in the literature, analyses which we show 
have by far underestimated the LQ contributions to 
the $\mu\to 3e$.
In particular we find that the coherent
$\mu-e$ conversion in nuclei 
mediated by the photon--conversion mechanism and the $\mu \to 3e$ decay
are golden plates where the flavor changing leptoquark couplings,
involving the second and third quark generations, can be strongly constrained.
This is due to the fact that these processes get the enhancements 
by large $\log(m_q^2/m^2_{LQ})$ terms which are induced by  
the so-called ``photon-penguin'' diagrams. These enhancements, 
which produce a mild GIM suppression in the amplitudes, 
have not been taken into account in the previous analyses.
We show that the $\mu \to e \gamma$ decay 
can set weaker constraints on the LQ models and this is because
its amplitude is strongly GIM suppressed by the 
terms of order $O(m_q^2/m^2_{LQ})$.
We also present the results for the corresponding constraints 
in the $\tau$ sector.
Finally the prospects of the future muon experiments for the
improvement of the present bounds are analyzed and discussed.
\vskip 11pt 
\vfil

\vskip 22pt 
\vfil
\noindent
{\it Electronic address}: emidio.gabrielli@cern.ch,~
emidio.gabrielli@uam.es
\end{titlepage}
\section{Introduction}
The leptoquark (LQ) color triplet bosons, predicted 
by the grand unified theories \cite{GUT}, superstrings 
inspired $E_6$ models \cite{string1}-\cite{string2},
the compositeness \cite{compos}, and the technicolor models \cite{tecni}
have been the subject of numerous phenomenological investigations.
The main direct experimental searches for LQs investigate
their production in the $s$-channel \cite{string1}, \cite{LQmodel}, 
these searches are carried out at the
{\it e--p} collider HERA at DESY \cite{LQmodel}-\cite{LQHERA}. 
On the other hand, the indirect searches of LQs consist mainly in analyzing  
the anomalous effects induced by the LQ interactions 
in the deep inelastic scatterings as well as in the low energy processes
\cite{Leptonrare}-\cite{LQrare}.\\
Recently there has been a renewed interest in the subject \cite{LQphen}
due to the high $Q^2$ anomalous events observed in the H1 \cite{H1} and 
ZEUS \cite{ZEUS} experiments at HERA, although subsequent 
analyses of new data have shown a less significant discrepancy with the 
standard model (SM) predictions \cite{H1new}. 

In the literature \cite{GUT}-\cite{compos} 
predictions are given for the existence of two types of LQs:
the scalar and vectorial ones.
In the present article we perform a model-independent study of the
LQs, and, in particular, we restrict ourselves to the most general
renormalizable interactions, 
which conserve the barion ($B$)
and lepton ($L$) number, and which are compatible with the
SM symmetries \cite{LQmodel}.
Therefore, in the case of vectorial
LQs, we consider only the gauge-vectorial ones.
From a theoretical point of view there is no reason to believe that the
quark-lepton couplings, mediated by LQ interactions,
are simultaneously diagonalizable with the quark and lepton mass matrices.
As a consequence, both scalar and vectorial LQs 
can generate, when integrated out, effective flavor changing (FC)
interactions between quarks and leptons of the form 
\cite{Leptonrare}-\cite{LQrare}
\be
\frac{\lambda_{LQ}^{n i}\lambda_{LQ}^{\dag j m}}{m_{LQ}^2}
\left(\bar{q}^i\gamma_{\mu} P^q q^j\right)
\left(\bar{l}^m\gamma_{\mu} P^l l^n\right),~~~
{\rm and/or},~~~ 
\frac{\lambda_{LQ}^{n i}\lambda_{LQ}^{\dag j m }}{m_{LQ}^2}
\left(\bar{q}^i P^q q^j\right)
\left(\bar{l}^m P^l l^n\right)
\ee
where $m_{LQ}$ indicates the LQ mass and 
the chiral projectors $P^{q}$ and $P^{l}$
of the quarks and lepton, resp., can be left ($P_L=(1-\gamma_5)/2$)
or right ($P_R=(1+\gamma_5)/2$).
Since these effective interactions can be generated even at tree-level,
strong constraints on the LQ 
couplings $\lambda_{LQ}^{ij}$ and masses can be set by the 
flavor changing neutral current (FCNC) processes.
In particular strong constraints on LQ models are obtained by 
means of the rare FC $K$, $D$, and $B$ meson decays 
\cite{LQrareL}-\cite{LQrare} as well as in the leptonic sector \cite{LQrare}.
In this framework in Ref.\cite{LQrare} a 
model-independent analysis was performed to constrain 
the masses and couplings of the LQs ($B$ and $L$ conserving).\\
In the present article we carry out a detailed analysis of 
the LQ contributions to the $\mu$ and $\tau$ leptonic rare processes.
Since we do not
consider the CP violating processes where 
the imaginary parts of $\lambda_{LQ}^{ij}$ couplings are better constrained,
we assume, as in Ref.\cite{LQrare},
that all the couplings $\lambda_{LQ}^{ij}$ are real.
In addition we require that all the couplings $\lambda_{LQ}^{ij}$ should be
unitary\footnote{
Note, however, that unitary LQ couplings can naturally appear
in some ``minimal'' models where the LQ interactions are assumed to be 
universal (in flavor) in the basis of the quark and lepton 
SM gauge--eigenstates.
This is for example the case of the gauge--vectorial LQs 
which appear in the gauge multiplet of the standard GUT theories \cite{GUT}.
Then, after rotating the quark and lepton fields
into the corresponding mass--eigenstates, 
unitary flavor--changing LQ couplings should
appear in the interactions in Eq.(\ref{LQlag})
being proportional to the products of 
the (unitary) diagonalization--matrices of the quark and lepton
mass matrices.} in both vector and scalar LQ sectors.
This last assumption will give us conservative limits.
Indeed the unitarity of the LQ couplings makes active
a GIM--type mechanism in some FCNC processes,
such as for example $\mu\to e\gamma$, $\tau\to e\gamma$ decay, 
$K-\bar{K}$, or $B-\bar{B}$ mixing, which naturally 
suppresses the potentially large LQ contributions.
We recall that the limits 
obtained in Ref.\cite{LQrare} 
(from $\mu\to e \gamma$ or $\tau\to e \gamma$)
in the scalar sector are less conservative 
than ours since the authors of Ref.\cite{LQrare} 
assumed unitary couplings only in the gauge-vectorial sector.

Within the class of interesting processes, used to constrain
the LQ models in the leptonic sector,
a special role is played by the rare FC violating decays 
$\mu \to e \gamma$ and $\mu^{-} \to e^{-}e^{-}e^{+}$ ($\mu \to 3e$)
\cite{LQrare}
and analogous processes in the $\tau$ sector, and the coherent
$\mu-e$ conversion in nuclei \cite{mue_old}-\cite{mue_huitu}
(in the following for the $\mu-e$ conversion we always 
mean the coherent process). 
The last of these processes is used to set the strongest constraints on the
combination $\lambda_{LQ}^{2 1}\lambda_{LQ}^{1 1}/m_{LQ}^2$
for both the vectorial and scalar LQs which involve the first
generation of quarks, since, in this case, the
effective Hamiltonian is induced at tree-level.
On the contrary the $\mu \to e \gamma$ and 
$\mu \to 3e$ decays, which are induced at one-loop,
allow to constrain the complementary combinations on
couplings and masses involving the second and third generation of quarks, 
namely
$\lambda_{LQ}^{2 i}\lambda_{LQ}^{1 i}/m_{LQ}^2$
or $\sqrt{\lambda_{LQ}^{2 i}\lambda_{LQ}^{1 i}}/m_{LQ}^2$, where $i=2,3$.
In this respect the authors of Ref.\cite{LQrare} found
that the radiative muon decay $\mu \to e \gamma$ is 
better than $\mu \to 3e$ process in setting stronger bounds on 
these combination of couplings. \\
One of the aims of this paper is to show that the conclusions
of Ref.\cite{LQrare} do not hold when the unitarity of the LQ couplings is
required in both scalar
and vectorial LQ sectors and the dominant
diagrams (not included in Ref.\cite{LQrare}) are taken into account 
in the $\mu \to 3e$ decay.
At 1-loop level the LQs give contributions to the $\mu \to e \gamma$ decay 
by means of the so called magnetic-penguin diagrams shown in Fig.[1d--e].
In Ref.\cite{LQrare} it is found that the $\mu \to e \gamma$ amplitude, 
mediated by 
scalar LQs (which couple to the external leptons with the same chirality), 
is enhanced with respect to the one corresponding to the 
gauge-vectorial LQs, since these last ones are strongly GIM suppressed
by terms of order $m_q^2/m_{LQ}^2$ 
(where $m_q$ is the typical quark mass running in the loop).
As a consequence they show that the scalar LQ
couplings and masses are more strongly constrained
than the corresponding vectorial ones.\\
In the present work we prove that, when the unitarity for the 
LQ couplings is extended to the scalar sector, the 
scalar LQ contribution to the $\mu \to e \gamma$ decay's
amplitude is significantly suppressed by the GIM mechanism
so that it turns out to be of the same order of the gauge-vectorial one.
This in particular implies that the constraints on scalar 
LQ couplings, which we find coming
from the $\mu \to e \gamma$ decay, are weaker than the corresponding ones given
in Ref.\cite{LQrare} 
and are roughly of the same order of the gauge-vectorial ones. 
\\
In addition we find that in Ref.\cite{LQrare} 
the main diagrams contribution to the 
$\mu \to 3e$ decay has been underestimated.
The main purpose of this paper, however, is to show that
the $\mu \to 3e$ process
is more powerful than the $\mu \to e\gamma $ one
in setting strong bounds on mass and couplings of scalar and
gauge-vectorial LQs. (We will see that similar
considerations hold for the analogous decays in the $\tau$ sector). 
This result can be simply understood as follows.
The main diagrams contribution to the $\mu \to 3e$
decay derives from the so called photon-penguins (see Fig.[1a--b])
which were not taken into account in Ref.\cite{LQrare}.
In the large LQ mass ($m_{LQ}$) limit (with respect to the quark mass
$m_q$) the photon-penguins, since they are proportional to $\log(m_q/m_{LQ})$,
are only ``mildly'' GIM suppressed
(instead of being ``strongly'' GIM suppressed as 
the box and Z-penguin diagrams contributions are).
This means that to have an extra electromagnetic coupling $\alpha$ in the 
$\mu \to 3e$ rate
(with respect to the $\mu \to e \gamma$ one) 
is a more convenient price to pay
than the price of the $O(m_q^4/m_{LQ}^4)$ suppression in 
the $\mu \to e \gamma$ rate in setting stronger bounds.
These $\log$ enhancements in $\mu \to 3e$ were known in 
literature \cite{log1} for certain models (and more recently 
analized in the context of effective theories \cite{log2}), 
however they were not applied in this context.
In fact the authors of Ref.\cite{LQrare}, in order to simplify the analysis, 
only consider the box--diagrams contribution
to the $\mu \to 3e$ decay (which are of the order 
$O(\lambda^4_{LQ} m_q^2/m_{LQ}^4)$) 
and this resulted in underestimating the branching ratio.\\

The photon-penguin diagrams also give a relevant contribution to
the $\mu-e$ conversion in nuclei \cite{mue_old}-\cite{mue_huitu} via
the photon--conversion mechanism.
The $\log$ enhancement of these diagrams 
will enable us to show that the $\mu-e$ conversion in nuclei 
is another golden-plate process which sets strong bounds for 
LQ couplings involving the second and third quark generations. 
These bounds could turn out to be more competitive 
than the corresponding ones obtained from the $\mu\to 3e$ process.
However we stress that the bounds from the $\mu-e$ conversion
suffer the problem of being model dependent due
to the non-perturbative calculations of the nuclear form factors,
while the bounds coming from the $\mu\to 3e$ decay are not.
\\
The interesting aspects of the photon-penguin diagrams, which
give universal contributions to the 
$\mu\to 3e$ and $\mu-e$ conversion processes,
is that these log enhancements
appear in both the scalar and vectorial LQ exchanges 
which have the same chirality couplings with the external leptons. 
We shall see that this property enables us to set strong bounds in both 
the scalar and vectorial LQ sectors.
\\
The same considerations regarding the $\log$ enhancements hold for the
$\tau \to 3e$, $\tau \to 3\mu$, $\tau \to e \mu^+\mu^-$, and 
$\tau \to \mu e^+e^-$ processes.
However, due to a weaker experimental upper limits on the
branching ratios, the constraints on the LQ couplings and masses,
which come from these rare processes, 
are not as strong as they are in the $\mu$ sector.

The paper is organized as follows.
In section [2]  we present the analytical results for the
LQ contributions (in the large LQ mass limit)
to the branching ratios of $\mu\to e \gamma$ and in section [3]
we give the corresponding results for  $\mu\to 3e$ decay 
and $\mu-e$ conversion.
In section [4] we present the numerical results for 
bounds on the combination of LQ couplings and masses 
constrained by the experimental upper limits on $\mu\to e\gamma$, 
$\mu\to 3e$ decay (and analogous processes in 
the $\tau$ sector), and the $\mu-e$ conversion.
Finally the last section is devoted to our conclusions.
\section{$\mu\to e \gamma$ decay}
In the present section we give the main results 
for the relevant LQ contributions to the total branching ratio of
$\mu\to e \gamma$.
We begin our analysis by fixing the conventions for
the most general renormalizable lagrangian
for scalar and gauge-vectorial LQ interactions (from now on, if not 
strictly necessary, we will omit the suffix gauge in the vectorial LQ).
This lagrangian, which is
$B$ and $L$ conserving and 
invariant under the $SU(3)_c\otimes SU(2)_L\otimes U(1)$
symmetry group of the SM,
was first proposed in Ref.\cite{LQmodel}. Its expression, in the
notation of Ref.\cite{LQrare}, is given by
\bea
{\cal L}_S&=&\left\{ \left(\lambda_{LS_0} \bar{q}^c_L\imath \tau_2 l_L
+\lambda_{RS_0} \bar{u}^c_R e_R\right)S_0^{\dag}
+ \left(\lambda_{R\tilde{S}_0} \bar{d}^c_R e_R\right) \tilde{S}_0^{\dag}
\right.\no
&+&\left.
\left(\lambda_{LS_{1/2}} \bar{u}_R l_L+
\lambda_{RS_{1/2}} \bar{q}_L \imath \tau_2 e_R\right) S^{\dag}_{1/2}
+
\left(\lambda_{L\tilde{S}_{1/2}} \bar{d}_R l_L\right) \tilde{S}_{1/2}^{\dag}
\right.\no
&+&\left.
\left(
\lambda_{LS_{1}} \bar{q}_L^c \imath \tau_2 \tau^{i} l_L\right)
S_1^{i}\right\} + h.c.,\no
{\cal L}_V&=&\left\{ \left(\lambda_{LV_0} \bar{q}_L\gamma_{\mu}l_L
+\lambda_{RV_0} \bar{d}_R \gamma_{\mu}e_R\right)V_0^{\dag \mu}
+ \left(\lambda_{R\tilde{V}_0} \bar{u}_R \gamma_{\mu} 
e_R\right) \tilde{V}_0^{\mu\dag}
\right.\no
&+&\left.
\left(\lambda_{LV_{1/2}} \bar{d}^c_R\gamma_{\mu} l_L+
\lambda_{RV_{1/2}} \bar{q}^c_L\gamma_{\mu}e_R\right) V^{\mu\dag}_{1/2}
+
\left(\lambda_{L\tilde{V}_{1/2}} \bar{u}^c_R\gamma_{\mu} 
l_L\right) \tilde{V}_{1/2}^{\mu\dag}
\right.\no
&+&\left.
\left(
\lambda_{LV_{1}} \bar{q}_L\gamma_{\mu}\tau^{i} l_L\right)
V^{\mu i \dag}_1\right\} + h.c.,
\label{LQlag}
\eea
where ${\cal L}_S$ and ${\cal L}_V$ contain 
the interactions with the scalar ($S_0,~\tilde{S}_0,~S_{1/2},~\tilde{S}_{1/2}
~S^i_{1}$)
and vectorial ($V_0^{\mu},~\tilde{V}_0^{\mu},~V_{1/2}^{\mu},~
\tilde{V}_{1/2}^{\mu},~V_{1}^{i \mu}$) LQs fields, respectively.
The subscript $(0,1/2,1)$ in each scalar and vectorial LQ 
indicates the 
{\it singlet}, 
{\it doublet}, and 
{\it triplet} $SU(2)_L$ representation, respectively, whereas
the $\tau^i$s are the Pauli matrices. 
The quark fields $q^c_{L,R}$ are the corresponding conjugate 
of the $q_{L,R}$ fields respectively, where $q_{L,R}^c\equiv (P_{L,R}q)^c$.
Note that the generation (flavor) and color
indices in the fields appearing in Eq.(\ref{LQlag}) are omitted. 
As we pointed out in the introduction, we assume
that all the scalar and vectorial LQ couplings $\lambda_{LQ}$ are unitary 
(in the flavor space) in the basis of the
quark and lepton mass eigenstates, and that the LQs 
do not carry flavor indices.
Moreover, since we do 
not consider the CP violating processes in our analysis, we assume that all the
couplings are real.

The relevant (gauge-invariant) effective Hamiltonian for the
$\mu\to e \gamma$ decay is given by
\be
H=\frac{4 G_F}{\sqrt{2}}\left(Q_{LR} C_{LR}+Q_{RL} C_{RL}\right),
\label{Hmurad}
\ee 
where $Q_{LR}=\bar{e}_L \sigma_{\mu\nu} \mu_R F^{\mu\nu}$
and $Q_{RL}=\bar{e}_R\sigma_{\mu\nu} \mu_L F^{\mu\nu}$
are the magnetic-dipole operators, $F^{\mu\nu}$ is the electromagnetic
field strength, and $C_{LR},~C_{RL}$ are the corresponding 
Wilson coefficients.\\
The Wilson coefficients $C_{LR},~C_{RL}$
receive their main contributions, at the electroweak scale,
through 1-loop magnetic penguin diagrams shown in Figs.[1d-e]. 
Since these diagrams are proportional 
to the $\sigma_{\mu\nu}$ form factor, one needs a chirality flip.
In the SM, as the charged currents
are only of the V-A type, one can get this chirality flip by means of an
external mass 
insertion.\footnote{In some extensions of the SM, such as 
the left--right--, the LQs--, and the supersymmetric--models,
(where the leptons or quarks can have 
both the left and right couplings to the new particles), 
this chirality flip can be realized by an heavy (internal) fermion mass
insertion which turns out 
to give a strong chiral enhancement respect to the SM amplitude.}
This imply that the $C_{LR}$ and $C_{RL}$ 
are proportional to the electron $m_e$ and muon $m_{\mu}$ masses, respectively.
However in the SM the Wilson coefficients $C_{LR}$ and $C_{RL}$
are strongly suppressed by the GIM mechanism which forces 
them to be proportional to
$m_{\nu}^2/m_W^2$ terms times the corresponding CKM angles 
of the leptonic sector, 
with $m_{\nu}$ being the heaviest neutrino mass running in the loop.\\
We now consider the LQ contributions to the magnetic penguin
diagrams (see Fig.[1d--e]). These diagrams receive finite contributions
from both the scalar and vectorial LQs interactions 
in Eq.(\ref{LQlag}). With 
respect to the SM diagrams, the W and neutrino internal lines
are replaced by a LQ and quark, respectively, and
an additional diagram (where the external photon is attached 
to the internal quark line) is included.
The SM GIM suppression terms of order $O(m_{\nu}^2/m_W^2)$ are 
now replaced by $O(m_{q}^2/m_{LQ}^2)$ where $m_{LQ}$ is a typical LQ 
mass running in the loop.\\
In order to find the constraints on the combination of 
LQ couplings and masses we impose, as in Ref.\cite{LQrare},
that each individual LQ coupling contribution to the branching ratio
does not exceed (in absolute value) the experimental upper limit on the
branching ratio.
Therefore, in order to simplify the analysis, we consider
in the branching ratio only one single LQ coupling contribution each time; 
this is done by ``switching off'' all the other couplings.
Moreover we assume that the LQ masses are larger than the
quark ones (including the case of the top quark) 
so that we only take the leading contribution to the Wilson coefficients. 

By means of the lagrangian in Eq.(\ref{LQlag}) we give in the sequel,
the results for the total branching ratio $\mu\to e\gamma$,
where only the contribution of a single LQ coupling 
is considered. By neglecting the terms of order $O(m_q^4/m_{LQ}^4)$ in the
Feynman diagrams we obtain 
\begin{itemize}
\item{Gauge-Vectorial LQs}
\bea 
B_R^{\lambda_{LV_0}}&=&\frac{B_V}{m_{V_0}^8G_F^2}
\left(Q_D-\frac{Q_{V_0}}{2}\right)^2
\left[
\sum_{i=1,3}\lambda_{LV_0}^{2i}\lambda_{LV_0}^{1i}
m_{D_i}^2\right]^2\no
B_R^{\lambda_{RV_0}}&=&B_R^{\lambda_{LV_0}}
(\lambda_{LV_0}\to \lambda_{RV_0})\no
B_R^{\lambda_{R\tilde{V}_{0}}}&=&
\frac{B_V}{m_{\tilde{V}_0^2}^8G_F^2}
\left(Q_U-\frac{Q_{\tilde{V}_0}}{2}\right)^2
\left[
\sum_{i=1,3}\lambda_{R\tilde{V}_0}^{2i}\lambda_{R\tilde{V}_0}^{1i}
m_{U_i}^2\right]^2\no
B_R^{\lambda_{LV_{1/2}}}&=&\frac{B_V}{m_{V_{1/2}}^8G_F^2}
\left(Q_D+\frac{Q_{V^D_{1/2}}}{2}\right)^2
\left[
\sum_{i=1,3}\lambda_{LV_{1/2}}^{2i}\lambda_{LV_{1/2}}^{1i}
m_{D_i}^2\right]^2\no
B_R^{\lambda_{RV_{1/2}}}&=&\frac{B_V}{m_{V_{1/2}}^8G_F^2}
\left[
\sum_{q=U,D}\sum_{i=1,3}\lambda_{RV_{1/2}}^{2i}\lambda_{RV_{1/2}}^{1i}
m_{q_i}^2\left(Q_q+\frac{Q_{V^q_{1/2}}}{2}\right)\right]^2\no
B_R^{\lambda_{L\tilde{V}_{1/2}}}&=&\frac{B_V}{m_{\tilde{V}_{1/2}}^8G_F^2}
\left(Q_U+\frac{Q_{\tilde{V}^D_{1/2}}}{2}\right)^2
\left[
\sum_{i=1,3}\lambda_{L\tilde{V}_{1/2}}^{2i}\lambda_{L\tilde{V}_{1/2}}^{1i}
m_{U_i}^2\right]^2\no
B_R^{\lambda_{LV_1}}&=&\frac{B_V}{m_{V_{1}}^8G_F^2}
\left[
\sum_{i=1,3}\lambda_{LV_1}^{2i}\lambda_{LV_1}^{1i}\left(
m_{D_i}^2\left(Q_D-\frac{Q_{V^3_{1}}}{2}\right)
+2m_{U_i}^2\left(Q_U-\frac{Q_{V^{1}_{1}}}{2}\right)\right)
\right]^2
\label{BRmuegV}
\eea
\end{itemize}
\begin{itemize}
\item{Scalar LQs}
\bea B_R^{\lambda_{LS_0}}&=&\frac{B_S}{m_{S_0}^8G_F^2}
\left[
\sum_{i=1,3}\lambda_{LS_0}^{2i}\lambda_{LS_0}^{1i}
m_{U_i}^2\left(
Q_U\rho(x_{\scriptscriptstyle{U_iS_0}})-
Q_{S_0}\right)
\right]^2\no
B_R^{\lambda_{RS_0}}&=&B_R^{\lambda_{LS_0}}(\lambda_{LS_0}\to \lambda_{RS_0})
\no
B_R^{\lambda_{R\tilde{S}_{0}}}&=&
\frac{B_S}{m_{\tilde{S}_0^2}^8G_F^2}
\left[
\sum_{i=1,3}\lambda_{R\tilde{S}_0}^{2i}\lambda_{R\tilde{S}_0}^{1i}
m_{D_i}^2
\left(Q_D\rho(x_{\scriptscriptstyle{D_i\tilde{S}_0}})-Q_{\tilde{S}_0}\right)
\right]^2\no
B_R^{\lambda_{LS_{1/2}}}&=&\frac{B_S}{m_{S_{1/2}}^8G_F^2}
\left[
\sum_{i=1,3}\lambda_{LS_{1/2}}^{2i}
\lambda_{LS_{1/2}}^{1i}
m_{U_i}^2
\left(Q_U
\rho(x_{\scriptscriptstyle{U_i{S}_{1/2}}})+Q_{S^D_{1/2}}\right)
\right]^2\no
B_R^{\lambda_{RS_{1/2}}}&=&\frac{B_S}{m_{S_{1/2}}^8G_F^2}
\left[
\sum_{i=1,3}\lambda_{RS_{1/2}}^{2i}\lambda_{RS_{1/2}}^{1i}\left(
m_{D_i}^2\left(Q_D\rho(x_{\scriptscriptstyle{D_i{S}_{1/2}}})
+Q_{S^U_{1/2}}\right) \right.\right.\no
&+&\left.\left. m_{U_i}^2\left(Q_U\rho(x_{\scriptscriptstyle{U_i{S}_{1/2}}})
+Q_{S^D_{1/2}}\right)\right)
\right]^2\no
B_R^{\lambda_{L\tilde{S}_{1/2}}}&=&\frac{B_S}{m_{\tilde{S}_{1/2}}^8G_F^2}
\left[
\sum_{i=1,3}\lambda_{L\tilde{S}_{1/2}}^{2i}\lambda_{L\tilde{S}_{1/2}}^{1i}
m_{D_i}^2
\left(Q_D\rho(x_{\scriptscriptstyle{D_i{\tilde{S}}_{1/2}}})
+Q_{\tilde{S}^D_{1/2}}\right)^2
\right]^2\no
B_R^{\lambda_{LS_1}}&=&\frac{B_S}{m_{S_{1}}^8G_F^2}
\left[
\sum_{i=1,3}\lambda_{LS_1}^{2i}\lambda_{LS_1}^{1i}\left(
m_{U_i}^2\left(Q_U\rho(x_{\scriptscriptstyle{U_i{S}_1}})
-Q_{S^3_{1}}\right) \right.\right.\no
&+&\left.\left. 2m_{D_i}^2\left(Q_D\rho(x_{\scriptscriptstyle{D_i{S}_1}})
-Q_{S^1_{1}}\right)\right)
\right]^2,
\label{BRmuegS}
\eea
\end{itemize}
where $B_V=3\alpha N_c^2/(64\pi)$ and $B_S=\alpha N_c^2/(192\pi)$.
$N_c=3$ is the number of colors and
the function $\rho(x)=(11+6\log{x})/2$, $x_{ab}=m^2_a/m^2_b$,
$Q_U=2/3$, and $Q_D=-1/3$.
For the values of the LQ charges $Q_{LQ}$ the reader is referred to
Table [\ref{LQcharge}]. 
Note that the leading term in $m_q^2/m_{V_0}^2$ expansions
in $B_R^{\lambda_{LV_0}}$, $B_R^{\lambda_{RV_0}}$
is zero due to the fact that $Q_D-Q_{V_0}/2=0$.
Therefore the non-zero
LQs contributions to these branching ratios is highly suppressed 
since they are of the order  $O(m_q^8/G_F^2m_{V_0}^{12})$.
\\
We stress that our findings for the analytical expressions in 
Eq.(\ref{BRmuegS}), due to different assumptions
in the scalar LQ couplings sector, is somewhat different 
to the corresponding one in Ref.\cite{LQrare}.
Indeed, in Ref.\cite{LQrare}, only the terms at the zero order 
in the quark mass expansion give the leading scalar LQ contribution 
to $\mu\to e \gamma$.
However in our case, when the sum over the quark generations is performed, 
these terms vanish
due to the unitarity of the $\lambda_{LQ}$ matrices and survive only the 
next-to-leading ones which are suppressed by terms of order 
$O(m_q^2/m_{LQ}^2)$.
We check that our results are consistent with the 
analogous calculations obtained, 
for example, from the charged Higgs and supersymmetric contributions 
to the $b\to s \gamma$ amplitude \cite{bbmr}.

Since the doublet LQs 
can have both couplings $\lambda_L$ and $\lambda_R$ (with left  
and right chiralities, respectively),
one can get a chiral enhancement in the $\mu\to e \gamma$ amplitude 
by flipping the chirality with an internal (heavy) quark mass insertion
in contrast with the external muon mass.
In this case the resulting amplitude has to be
proportional to $\lambda_L\lambda_R$.
However in the present paper
we do not analyze the constraints on this combination of couplings.
Indeed we follow the usual approach, 
(adopted also in Ref.\cite{LQrare}), in setting more 
conservative bounds and, more precisely, consider only the 
effects of the same kind of coupling constant per time while ``switching off'' 
all the others.

In the $\tau$ sector the corresponding analytical results
for the $\tau \to e\gamma$ and $\tau \to \mu\gamma$ decays
are simply obtained by making the following substitutions in the 
right hand sides (r.h.s.) 
of the branching ratios $B_R^{\lambda_{LQ}}(\mu\to e\gamma)$ 
in Eqs.(\ref{BRmuegV})--(\ref{BRmuegS})
\bea
B_R^{\lambda_{LQ}}(\tau \to e\gamma)&=& 
B_R^{exp}(\tau^{-}\to \mu^{-}\bar{\nu}_{\mu}\nu_{\tau})
\times B_R^{\lambda_{LQ}}(\mu \to e\gamma)\left\{\lambda^{2i}_{LQ}
\lambda^{1i}_{LQ}\to 
\lambda^{3i}_{LQ}\lambda^{1i}_{LQ}\right\},\no
B_R^{\lambda_{LQ}}(\tau \to \mu\gamma)&=& 
B_R^{exp}(\tau^{-}\to \mu^{-}\bar{\nu}_{\mu}\nu_{\tau})\times
B_R^{\lambda_{LQ}}(\mu \to e\gamma)\left\{\lambda^{2i}_{LQ}
\lambda^{1i}_{LQ}\to 
\lambda^{3i}_{LQ}\lambda^{2i}_{LQ}\right\},
\label{BRtaulg}
\eea
where the electron and muon masses 
have been neglected with respect to the $\tau$ mass.
The central value of the experimental branching ratio for
$\tau^{-}\to \mu^{-}\bar{\nu}_{\mu}\nu_{\tau}$ is
$B_R^{exp}(\tau^{-}\to \mu^{-}\bar{\nu}_{\mu}\nu_{\tau})=
17.37\%$ \cite{PDG}.
\section{$\mu\to 3e$ decay and $\mu-e$ conversion in nuclei} 
The most general lepton-family violating ($\Delta L=1$)
effective Hamiltonian which describes the amplitude of 
the $\mu\to 3e$ decay is 
\bea
H^{(\Delta L=1)}&=&\frac{4 G_F}{\sqrt{2}}\left\{Q_{LR} C_{LR}+Q_{RL} C_{RL}
+C_1\left(\bar{e}_R \mu_L\right)\left(\bar{e}_R e_L\right)\right.\no
&+&\left.C_2\left(\bar{e}_L \mu_R\right)\left(\bar{e}_L e_R\right)
+C_3\left(\bar{e}_R \gamma_{\mu} \mu_R\right)
\left(\bar{e}_R \gamma^{\mu} e_R\right)
+C_4\left(\bar{e}_L \gamma_{\mu} \mu_L\right)
\left(\bar{e}_L \gamma^{\mu} e_L\right)
\right.\no
&+&\left. 
C_5\left(\bar{e}_R \gamma_{\mu} \mu_R\right)
\left(\bar{e}_L \gamma^{\mu} e_L\right)
+C_6\left(\bar{e}_L \gamma_{\mu} \mu_L\right)
\left(\bar{e}_R \gamma^{\mu} e_R\right)+h.c.
\right\},
\label{Hgen}
\eea
where $Q_{LR}$ and $Q_{RL}$ are the magnetic-dipole operators defined in
Sec.[2] and the chiral fields are given by $\psi_{L,R}=
\frac{1}{2}(1\mp \gamma_5)\psi$.
In the LQ model obtained with the Lagrangian (\ref{LQlag}),
the Wilson coefficients of the local four-fermion operators in 
(\ref{Hgen}) get their contributions from 
the photon-penguin, Z-penguins, and the box diagrams
(see Fig.[1a--c] with $f=e$ and scalar-- and vectorial--LQs 
running in the loops together with quarks).\footnote{
In the case of unitary couplings any single diagram of the gamma- or
Z-penguin type is finite. This is because the divergent part, 
being flavor independent, factorizes out in sum
$\sum_{i=1,3}\lambda_{LQ}^{\dag m i}\lambda_{LQ}^{i n}$ 
on the internal--flavor index $i$ and therefore 
its contribution vanishes for $m\neq n$.
In the case of non-unitary couplings, due 
to the $U(1)\times SU(2)_L$ Ward identities, only the total sum of the
gamma-penguin-diagrams, or analogously the Z-penguin ones
(including also the diagrams with the self-energy insertions),
are finite. 
On the contrary the magnetic-penguin- and box-diagrams 
(see Figs.[1c--d] respectively) 
are just finite since their loop integrals are convergent.
Clearly, in the vectorial sector, the above arguments hold
if the vectorial LQs are of the gauge-type and acquire mass via 
spontaneous symmetry breaking.}
In the limit of large LQ masses (large compared to the quark ones)
the leading contribution to the effective Hamiltonian
in (\ref{Hgen}) is given by the photon-penguin diagrams which affects
only the coefficients $C_3,~C_4,~C_5$, and $C_6$.
Indeed, in the large LQ mass limit, these diagrams are only
``mildly'' GIM suppressed since they are proportional 
to the $\lambda_{LQ}^2\log(m_q^2/m_{LQ}^2)/m_{LQ}^2$; 
this, in the context of the GIM mechanism, is
in contrast with the naive expectation
of them being of order $O(\lambda_{LQ}^2 m_q^2/m_{LQ}^4)$ and 
$O(\lambda_{LQ}^4 m_q^2/m_{LQ}^4)$ which is typical of
the Z-penguins and box diagrams respectively.
We stress that the appearance of the $\log(m_q^2/m_{LQ}^2)$ enhancements 
is a special property of the gamma-penguin diagrams which is 
independent by any assumption on the unitarity of the LQ couplings.

The relevance of these log enhancements 
in constraining the new physics beyond the SM was first analyzed in 
Ref.\cite{log1}.
A more accurate discussion on the origin of these logarithms can
be found in Ref.\cite{log2}.
Here we want to point out that this mild GIM suppression of 
the photon-penguins, which is also present in the SM, 
is a peculiar property of the LQ interactions and 
it is not true in general extensions of the SM. 
For example, in the minimal supersymmetric SM, due to the super-GIM
mechanism, the photon-penguins mediated by the charged Higgs, 
gauginos or Higgsinos are always GIM suppressed by terms of 
order $O(m_q^2/m_{SUSY}^2)$, 
(with $m_{SUSY}$ standing for any typical SUSY mass running in the loops)
\cite{bbmr}, \cite{gg}.
\\
Therefore, due to these log enhancements, the $\mu\to 3e$ decay
plays a crucial role in setting strong constraints on the 
FC LQs couplings in the $\Delta L=1$ processes.
The branching ratio for this process is proportional to 
$\alpha^2/(m_{LQ}^2 G_F)^2 \times \log(m_q/m_{LQ})$, 
while the branching ratio of the $\mu\to e\gamma$ decay is proportional to
$\alpha /(m_{LQ}^2 G_F)^2 \times (m_q/m_{LQ})^4$.
Therefore even though in the branching ratio of the $\mu \to 3e$ decay 
we pay the price of an extra electromagnetic $\alpha$ coupling,
with respect to the $\mu\to e\gamma$, in the large LQ
mass limit the $\alpha$ suppression could dominate 
the $m_q^4/m_{LQ}^4$ term.
For this reason we neglect the effects of the magnetic-penguins operators
$Q_{LR}$ and $Q_{RL}$ in the $\mu \to 3e$ branching ratio.\footnote{
We recall that if the Wilson coefficients of 
the magnetic-penguin operators in (\ref{Hgen})
are not suppressed then
their contribution can give sizeable effects to the $\mu\to 3e$ 
decay \cite{mu3eGUT}.
Moreover, after the phase space integration, some terms (proportional to 
the magnetic-penguin contributions) get an enhancement of  
$\log{(m_{\mu}/m_e)}$ in the decay rate because of
the $1/q^2$ pole of the gamma propagator.
Therefore, if these diagrams are included then the approximation consisting in
the neglect of the electron mass, cannot be used.}
The main contribution to the branching ratio of $\mu\to 3e$ is given by
\cite{mu3eGUT}
\be
B_R=2\left(\frac{|C_1|^2}{16}+\frac{|C_2|^2}{16}+
|C_3|^2+|C_4|^2\right)+|C_5|^2+|C_6|^2,
\label{BR}
\ee
where $C_i$ are defined in Eq.(\ref{Hgen}). Note that
the terms suppressed by $m_e/m_{\mu}$ have been neglected.
In addition in obtaining Eq.(\ref{BR}) 
the total width of the muon was approximated and we used instead the
width of the main decay $\mu^{-}\to \nu_{\mu} e^{-} \bar{\nu}_e$
whose branching ratio is almost $100\%$.
Moreover, by following the approach adopted in this analysis,
we consider the photon-penguins contributions
induced by only one LQ couplings and by ``switching off'' all the others.
By means of these approximations and by neglecting 
the terms proportional to the quark masses, the expression for the 
branching ratio in Eq.(\ref{BR}) can be further simplified, 
since in this case we have $C_{1,2}=0$, $C_3=C_5$ and $C_4=C_6$.
We next give the analytical results relative to the branching ratio of
$\mu \to 3e$ mediated by the photon-penguins;
(the analogous results in the $\tau$ sector can be obtained by means of a simple
generalization):
\begin{itemize}
\item{Gauge-Vectorial LQs}
\bea 
B_R^{\lambda_{LV_0}}&=&\frac{\tilde{B}_V}{m_{V_0}^4G_F^2}Q_D^2
\left[
\sum_{i=1,3}\lambda_{LV_0}^{2i}\lambda_{LV_0}^{1i}
\log(\frac{m_{D_i}^2}{m_{V_0}^2})\right]^2\no
B_R^{\lambda_{RV_0}}&=&B_R^{\lambda_{LV_0}}
(\lambda_{LV_0}\to \lambda_{RV_0})\no
B_R^{\lambda_{R\tilde{V}_{0}}}&=&
\frac{\tilde{B}_V}{m_{\tilde{V}_0^2}^4G_F^2}Q_U^2
\left[
\sum_{i=1,3}\lambda_{R\tilde{V}_0}^{2i}\lambda_{R\tilde{V}_0}^{1i}
\log(\frac{m_{U_i}^2}{m_{\tilde{V}_0}^2})\left(1+\delta_{i3}
\Delta_V(x_t,R_{\tilde{V}_0})\right)\right]^2\no
B_R^{\lambda_{LV_{1/2}}}&=&\frac{\tilde{B}_V}{m_{V_{1/2}}^4G_F^2}Q_D^2
\left[
\sum_{i=1,3}\lambda_{LV_{1/2}}^{2i}\lambda_{LV_{1/2}}^{1i}
\log(\frac{m_{D_i}^2}{m_{V_{1/2}}^2})\right]^2\no
B_R^{\lambda_{RV_{1/2}}}&=&\frac{\tilde{B}_V}{m_{V_{1/2}}^4G_F^2}
\left[
\sum_{i=1,3}\lambda_{RV_{1/2}}^{2i}\lambda_{RV_{1/2}}^{1i}\left(
Q_D\log(\frac{m_{D_i}^2}{m_{V_{1/2}}^2})\right.\right.\no
&+&\left.\left.Q_U\log(\frac{m_{U_i}^2}{m_{V_{1/2}}^2})\left(1+\delta_{i3}
\Delta_V(x_t,-R_{V^U_{1/2}})\right)\right)
\right]^2\no
B_R^{\lambda_{L\tilde{V}_{1/2}}}&=&
\frac{\tilde{B}_V}{m_{\tilde{V}_{1/2}}^4G_F^2}Q_U^2
\left[
\sum_{i=1,3}\lambda_{L\tilde{V}_{1/2}}^{2i}\lambda_{L\tilde{V}_{1/2}}^{1i}
\log(\frac{m_{U_i}^2}{m_{\tilde{V}_{1/2}}^2})\left(1+\delta_{i3}
\Delta_V(x_t,-R_{\tilde{V}^D_{1/2}})\right)
\right]^2\no
B_R^{\lambda_{LV_1}}&=&\frac{\tilde{B}_V}{m_{V_{1}}^4G_F^2}
\left[
\sum_{i=1,3}\lambda_{LV_1}^{2i}\lambda_{LV_1}^{1i}\left(
Q_D\log(\frac{m_{D_i}^2}{m_{V_{1}}^2})\right.\right.\no
&+&\left.\left.2Q_U\log(\frac{m_{U_i}^2}{m_{V_{1}}^2})\left(1+\delta_{i3}
\Delta_V(x_t,R_{V_1})\right)
\right)\right]^2
\label{BRmu3eV}
\eea
\end{itemize}
\begin{itemize}
\item{Scalar LQs}
\bea 
B_R^{\lambda_{LS_0}}&=&\frac{\tilde{B}_S}{m_{S_0}^4G_F^2}Q_U^2
\left[
\sum_{i=1,3}\lambda_{LS_0}^{2i}\lambda_{LS_0}^{1i}
\log(\frac{m_{U_i}^2}{m_{S_0}^2})\left(1+\delta_{i3}
\Delta_S(x_t,-R_{S_0})\right)
\right]^2\no
B_R^{\lambda_{RS_0}}&=&B_R^{\lambda_{LS_0}}
(\lambda_{LS_0}\to \lambda_{RS_0})\no
B_R^{\lambda_{R\tilde{S}_{0}}}&=&
\frac{\tilde{B}_S}{m_{\tilde{S}_0^2}^4G_F^2}Q_D^2
\left[
\sum_{i=1,3}\lambda_{R\tilde{S}_0}^{2i}\lambda_{R\tilde{S}_0}^{1i}
\log(\frac{m_{D_i}^2}{m_{\tilde{S}_0}^2})\right]^2\no
B_R^{\lambda_{LS_{1/2}}}&=&\frac{\tilde{B}_S}{m_{S_{1/2}}^4G_F^2}Q_U^2
\left[
\sum_{i=1,3}\lambda_{LS_{1/2}}^{2i}\lambda_{LS_{1/2}}^{1i}
\log(\frac{m_{U_i}^2}{m_{S_{1/2}}^2})\left(1+\delta_{i3}
\Delta_S(x_t,R_{S^D_{1/2}})\right)
\right]^2\no
B_R^{\lambda_{RS_{1/2}}}&=&\frac{\tilde{B}_S}{m_{S_{1/2}}^4G_F^2}
\left[
\sum_{i=1,3}\lambda_{RS_{1/2}}^{2i}\lambda_{RS_{1/2}}^{1i}\left(
Q_D\log(\frac{m_{D_i}^2}{m_{S_{1/2}}^2})\right.\right.\no
&+&\left.\left.
Q_U\log(\frac{m_{U_i}^2}{m_{S_{1/2}}^2})\left(1+\delta_{i3}
\Delta_S(x_t,R_{S^D_{1/2}})\right)\right)
\right]^2\no
B_R^{\lambda_{L\tilde{S}_{1/2}}}&=&
\frac{\tilde{B}_S}{m_{\tilde{S}_{1/2}}^4G_F^2}Q_D^2
\left[
\sum_{i=1,3}\lambda_{L\tilde{S}_{1/2}}^{2i}\lambda_{L\tilde{S}_{1/2}}^{1i}
\log(\frac{m_{D_i}^2}{m_{\tilde{S}_{1/2}}^2})\right]^2\no
B_R^{\lambda_{LS_1}}&=&\frac{\tilde{B}_S}{m_{S_{1}}^4G_F^2}
\left[
\sum_{i=1,3}\lambda_{LS_1}^{2i}\lambda_{LS_1}^{1i}\left(
Q_U\log(\frac{m_{U_i}^2}{m_{S_{1}}^2})\left(1+\delta_{i3}
\Delta_S(x_t,-R_{S_3})\right)\right.\right.\no
&+&\left.\left.2Q_D\log(\frac{m_{D_i}^2}{m_{S_{1}}^2})\right)\right]^2
\label{BRmu3eS}
\eea
\end{itemize}
where $\tilde{B}_V=\alpha^2 N_c^2/(96\pi^2)$ and 
$\tilde{B}_S=\alpha^2 N_c^2/(384\pi^2)$.
The functions
$\Delta_{S,V}(x_t,R_{LQ})$ (where $\delta_{ij}$ is the
standard delta function, $x_t=m_t^2/m^2_{LQ}$ and $R_{LQ}\equiv Q_{LQ}/Q_{U}$) 
whose expressions are given in appendix, take into
account the exact dependence on the top mass $m_t$. Note that in the limit
$\lim_{x_t\to 0}\left(\Delta_{S,V}(x_t,R_{LQ})\right)\to 0$.
The results in Eqs.~(\ref{BRmu3eV})--(\ref{BRmu3eS})
are in agreement with the analogous ones 
obtained in Refs.\cite{bbmr}, \cite{gg} for the SUSY corrections to the
FCNC processes in the quark sector.
\\
In the $\tau$ sector the corresponding analytical results
for the $\tau \to 3e$ and $\tau \to 3\mu$ decays
are obtained by making the following substitutions in the 
r.h.s. of Eqs.(\ref{BRmu3eV})--(\ref{BRmu3eS})
\bea
B_R^{\lambda_{LQ}}(\tau \to 3e)&=& 
B_R^{exp}(\tau^{-}\to \mu^{-}\bar{\nu}_{\mu}\nu_{\tau})
\times B_R^{\lambda_{LQ}}(\mu \to 3e)\left\{\lambda^{2i}_{LQ}
\lambda^{1i}_{LQ}\to 
\lambda^{3i}_{LQ}\lambda^{1i}_{LQ}\right\},\no
B_R^{\lambda_{LQ}}(\tau \to 3\mu)&=& 
B_R^{exp}(\tau^{-}\to \mu^{-}\bar{\nu}_{\mu}\nu_{\tau})\times
B_R^{\lambda_{LQ}}(\mu \to 3e)\left\{\lambda^{2i}_{LQ}\lambda^{1i}_{LQ}\to 
\lambda^{3i}_{LQ}\lambda^{2i}_{LQ}\right\}.
\label{BRtau3l}
\eea

We now consider the LQ contribution to the 
$\mu-e$ conversion in nuclei \cite{mue_old}-\cite{mue_huitu}. 
The most general effective Hamiltonian which is relevant for this process 
is given by
\bea
H^{(\Delta L=1)}_{hadr}
&=&\frac{4 G_F}{\sqrt{2}}\left\{Q_{LR} C_{LR}+Q_{RL} C_{RL}\right\}
+C^h_1\left(\bar{e}_R \mu_L\right)\sum_{q=u,d}
\left(\bar{q}_R q_L\right)\no
&+&C^h_2\left(\bar{e}_L \mu_R\right)
\sum_{q=u,d}\left(\bar{q}_L q_R\right)+
C^h_3\left(\bar{e}_L \mu_R\right)
\sum_{q=u,d}\left(\bar{q}_L q_R\right)+
C^h_4\left(\bar{e}_L \mu_R\right)
\sum_{q=u,d}\left(\bar{q}_L q_R\right)\no
&+&C^h_5\left(\bar{e}_R \gamma_{\mu} \mu_R\right)
\sum_{q=u,d}Q_q\left(\bar{q}_R \gamma^{\mu} q_R\right)
+C^h_6\left(\bar{e}_L \gamma_{\mu} \mu_L\right)
\sum_{q=u,d}Q_q\left(\bar{q}_L \gamma^{\mu} q_L\right)\no
&+&
C^h_7\left(\bar{e}_R \gamma_{\mu} \mu_R\right)
\sum_{q=u,d}Q_q\left(\bar{q}_L \gamma^{\mu} q_L\right)
+C^h_8\left(\bar{e}_L \gamma_{\mu} \mu_L\right)
\sum_{q=u,d}Q_q\left(\bar{q}_R \gamma^{\mu} q_R\right)+h.c.~,
\label{Hmue}
\eea
where in the quark $q$ fields the sum over color indices is assumed and
$Q_q$ is the quark electric charge in the unit of $e$. Note that
for the Wilson coefficients $C_i^h$ we have used a normalization 
which is different from the one used in (\ref{Hgen}).
In the LQ model the Wilson coefficients of the 
four fermion operators get their contribution at tree level with a LQ
exchange. The tree-level contributions involve only the products of the
LQ couplings 
$\lambda_{LQ}^{2 1}\lambda_{LQ}^{1 1}$, since the operators in 
(\ref{Hmue}) contain only quarks of the first generation.
Indeed this process is known to be particularly effective 
for setting strong bounds on this combination of couplings
\cite{Leptonrare}, \cite{LQrare}.
However the Wilson coefficients $C_i^h$ also 
receive the next-to-leading 
contributions from the 1-loop diagrams 
induced by the photon-penguins, Z-penguins, and box diagrams
(see Fig.[1a--c] with $f=U,D$ quarks). 
These next-to-leading contributions involve the products of the 
$\lambda_{LQ}^{2 i}\lambda_{LQ}^{1 i}$ combinations with $i=1,2,3$; as a result
the $\mu-e$ conversion process can also give the possibility 
to set constraints on the second and third quark generations.\footnote{
In a more recent paper the authors of Ref. \cite{mue_huitu}
stress the importance of the log enhancements, induced
in the  
$\mu-e$ conversion rate by the photon-penguins, in constraining 
the supersymmetric R-parity violating models. However in these models 
the tree-level contributions to the $\mu-e$ conversion are absent.}
In particular, in order to set bounds from the $\mu-e$ conversion on 
the combinations $\lambda_{LQ}^{2 i}\lambda_{LQ}^{1 i}/m_{LQ}^2$ 
with $i=2,3$, we assume that its branching ratio is dominated by
the 1-loop contributions. This implies that the product
$\lambda_{LQ}^{2 1}\lambda_{LQ}^{1 1}$ should be negligible with respect to
the products $\lambda_{LQ}^{2 i}\lambda_{LQ}^{1 i}$ (with $i=2,3$)
times electromagnetic constant $\alpha$.
This condition does not violate the unitarity of the $\lambda_{LQ}$
matrices since in this case one can still have
$\lambda_{LQ}^{2 2}\lambda_{LQ}^{1 2}=-\lambda_{LQ}^{2 3}\lambda_{LQ}^{1 3}
+O(\lambda_{LQ}^{2 1}\lambda_{LQ}^{1 1})$.
Now, even though the constraints on couplings involving the second and third 
quark generations are weaker than the corresponding ones with the
first generation, they should be compared to the same bounds obtained
by other processes, such as for example the $\mu\to 3e$ decay.
Indeed, as we will show in the next section, the bounds on the LQ couplings
involving the second and third quark generations, which obtained from
the current experimental upper limit on the $\mu-e$ conversion rate,
are stronger than the corresponding ones obtained from the $\mu\to 3e$ decay.

In the large LQ mass limit, the photon-penguin diagrams, 
due to the log enhancements, dominate the other 1-loop diagrams, 
such as the Z-penguins and box diagrams which are of order $O(m_q^2/m_{LQ}^2)$.
Therefore in constraining the LQ couplings involving the
second and third quark generations we assume that the $\mu-e$ conversion
is dominated by the photon-conversion mechanism
\cite{mue_old}-\cite{mue_kosmas}. In this case
only the matrix elements of the
hadronic electromagnetic current between nuclei are involved.
In order to calculate the nuclear form factors 
connected to the electromagnetic matrix elements
various models and approximations have been applied in literature
\cite{mue_old}-\cite{mue_kosmas}. In Ref.\cite{mue_shank}, and more recently
in Ref.\cite{marciano}, the relativistic
effects have been taken into account.
However, due to the large mass of the muon, it is useful to 
take the non-relativistic limit 
of the motion of the muon in the muonic atom (see Ref.\cite{mue_chiang}
and more recently Ref.\cite{mue_kosmas}).
Indeed, in this limit, the large uncertainty
connected to the muon wave function factorizes out in the calculation of
the coherent conversion rate.
In order to estimate the LQ contribution to the  
$\mu-e$ conversion in nuclei $N$ 
(defined as $B_R(\mu-e)_N\equiv \Gamma(\mu~N \to e~N)/
\Gamma(\mu~N \to \nu_{\mu}~N^{\prime})$),
we use the non-relativistic results of Ref.\cite{mue_chiang}.
In the photon-conversion mechanism 
one can set $C_i^h=0,~~i=1,\dots,4$, $C_5^h=C_7^h$, and $C_6^h=C_8^h$ in the
Hamiltonian (\ref{Hmue}), and the branching ratio is given by 
\cite{mue_chiang}, \cite{mue_huitu}
\be
B_R(\mu-e)_N=
C\frac{\alpha^3 m_{\mu}^5 Z_{eff}^4 Z |\bar{F}_p|^2}
{\Gamma_{capt}} \frac{1}{4\pi^2}\left(|C^h_5|^2+|C_8^h|^2\right),
\label{rate}
\ee
where $\bar{F}_p(q)$ is the proton nuclear form factor 
(see Ref.\cite{mue_chiang} for more details) and $\Gamma_{capt}$
is the total muon capture rate.\footnote{
Note that with respect to the corresponding formula of Ref.\cite{mue_huitu}, 
we have simplified the $1/q^2$ pole of the
photon propagator (which for this process is of order $|q^2|\simeq m_{\mu}^2$)
with the $q^2$ which, due to the gauge-invariance, 
factorizes in the photon-penguin form factor.}
All these quantities depend upon the Titanium element 
($_{22}^{48}Ti$) used in the current experiment \cite{mueEXP}. 
For the $Ti$ the quantities appearing in Eq.(\ref{rate})
take the following values \cite{mue_chiang}: $C^{Ti}=1.0$,
$Z_{eff}^{Ti}=17.61$, $Z^{Ti}=22$, $\Gamma_{capt}^{Ti}=2.59\times 
10^6~s^{-1}$, and $\bar{F}_p^{Ti}(q)=0.55$.
\\
By taking into account only the photon-penguin contributions we see that, 
respectively, the $C_5^h$ and $C_6^h$ coefficients
are proportional to $C_3$ and $C_6$ in Eq.({\ref{Hgen}) by
an overall factor. This factor is a constant and
does not depend upon the particular LQ model considered.
This implies that the LQ contributions to the 
branching ratio of the $\mu-e$ conversion in nuclei, mediated by the
photon-conversion mechanism, have the same expression 
of the Eqs.(\ref{BRmu3eV})--(\ref{BRmu3eS}) and so the bounds extracted from 
this process can be simply obtained by rescaling 
the relative bounds obtained from the $\mu\to 3e$ decay.
In order to obtain the corresponding analytical expressions of 
Eqs.(\ref{BRmu3eV})--(\ref{BRmu3eS}) 
for the branching ratios of the $\mu-e$ photon--conversion in nuclei
the following substitution has to be made in the r.h.s of
Eqs.(\ref{BRmu3eV})--(\ref{BRmu3eS})
\be
B_R^{\lambda_{LQ}}(\mu-e)_{Ti}=B_R^{\lambda_{LQ}}(\mu \to 3e)
\left\{\frac{1}{G_F^2} \to C\frac{\alpha^3 m_{\mu}^5 Z_{eff}^4 Z |\bar{F}_p|^2}
{\Gamma_{capt}} \frac{2}{3\pi^2}\right\},
\label{rescaling}
\ee
where, as in the $\mu\to 3e$ case, we make active the effect of one 
single coupling $\lambda_{LQ}$ per time by ``switching off'' all the others.
\section{Numerical results for the bounds}
In this section we present the numerical results for the bounds on the
combinations of LQ couplings $\lambda_{LQ}^{ij}$ and masses $m_{LQ}$ 
obtained from the $\mu\to 3e$, $\mu-e$ conversion in $Ti$, and 
the $\mu\to e\gamma$.
In addition we give the results for the corresponding bounds 
in the $\tau$ sector.
In setting constraints
on the product of couplings 
$\lambda_{LQ}^{2 i}\lambda_{LQ}^{1 i}$ ($i=1,2,3$), in order to simplify 
our study, we use the following approach
\begin{itemize}
\item The tree-level LQ's contribution to the 
$\mu-e$ conversion rate (not mediated by 
the photon--conversion mechanism) contains only the terms
$\lambda_{LQ}^{21}\lambda_{LQ}^{11}$.
This process is used to strongly
constrain the combinations $|\lambda_{LQ}^{21}\lambda_{LQ}^{11}/m_{LQ}^2|$ .
\item
In order to set bounds on the LQ couplings involving the 
second and third generations of quarks, we adopt the approximation
which neglects $\lambda_{LQ}^{21}\lambda_{LQ}^{11}$ with respect to
$\lambda_{LQ}^{22}\lambda_{LQ}^{12}$ and $\lambda_{LQ}^{23}\lambda_{LQ}^{13}$
in the one-loop contributions.
This approximation can be justified as follows.
The products of couplings $\lambda_{LQ}^{2i}\lambda_{LQ}^{1i}$, $i=2,3$
enter only in the
one-loop contribution to the $\mu\to e \gamma$, $\mu\to 3e$ decays, 
and the $\mu-e$ conversion.
Therefore they 
could be larger if compared with the $\lambda_{LQ}^{21}\lambda_{LQ}^{11}$ ones
without violating the experimental upper limits on the branching ratios.
\item
In this approximation the unitarity of the
$\lambda_{LQ}$ matrices implies that:
$\lambda_{LQ}^{2 2}\lambda_{LQ}^{1 2}=-\lambda_{LQ}^{2 3}\lambda_{LQ}^{1 3}$.
This condition allows us to eliminate one of the two products of couplings 
in the one-loop contributions
and set bounds for the magnitude of the combinations
$|\lambda_{LQ}^{2 2}\lambda_{LQ}^{1 2}|/m_{LQ}^2$ or
$|\lambda_{LQ}^{2 3}\lambda_{LQ}^{1 3}|/m_{LQ}^2$ by means of 
Eqs.(\ref{BRmuegV})-(\ref{BRmuegS}), (\ref{BRmu3eV})-(\ref{BRmu3eS}).
In order to set bounds in the sector $i$, with $i=2,3$, by means of the 
$\mu-e$ conversion one assumes that (for this process) the photon-conversion
mechanism dominates over the non-photonic one.
\item
The same approach is adopted in setting bounds in the $\tau$ sector.
The relevant effective Hamiltonians for the 
$\tau\to \pi^0 e$ and $\tau\to \pi^0 \mu$ decays are induced at the tree-level
and so these processes are used to strongly constrain the variables
$|\lambda_{LQ}^{31}\lambda_{LQ}^{11}|/m_{LQ}^2$
and $|\lambda_{LQ}^{31}\lambda_{LQ}^{21}|/m_{LQ}^2$, respectively. 
The terms 
$\lambda_{LQ}^{3i}\lambda_{LQ}^{1i}$ and $\lambda_{LQ}^{3i}\lambda_{LQ}^{2i}$
(involving the second and third quark generations) enter at one-loop
level and therefore, as in the muon sector, we assume that they could be 
larger than the corresponding ones involving the first generation.
The bounds on these combinations of couplings
are obtained by imposing the experimental upper limits on the  
$\tau\to e\gamma$, $\tau\to 3e$, $\tau\to \mu\gamma$, 
and $\tau\to 3\mu$ decays.
\end{itemize}

In tables [\ref{tabS}]-[\ref{tabDV}]
we present our results for the upper bounds on the following variables
$\xi^{i}_{LQ}\equiv |\lambda_{LQ}^{2 i}\lambda_{LQ}^{1 i}|\times
(10^2 {\rm GeV}/m_{LQ})^2$ and 
$\tilde{\xi}^{i}_{LQ}\equiv \sqrt{|\lambda_{LQ}^{2 i}\lambda_{LQ}^{1 i}|}\times
(10^2 {\rm GeV}/m_{LQ})^2$
with $i=1,2,3$.
The constraints on $\xi^1_{LQ}$ are obtained by using the same approach 
used in Ref.\cite{LQrare}.
In addition, by means of 
the current experimental upper limits on the $\mu-e$ conversion rate 
on $Ti$ which is $B_R^{exp}(\mu-e)_{Ti}<6.1\times 10^{-13}$ \cite{mueEXP},
we improve the results of Ref.\cite{LQrare}.
From tables [\ref{tabS}]-[\ref{tabDV}] we see 
that the current limits on the 
$\mu-e$ conversion rate can set bounds on $\xi^1_{LQ}$ which are at the level
of $\xi^1_{LQ} < O(10^{-7})$.
Our results for the bounds on  $\xi_{LQ}^1$ are in agreement with the
corresponding ones in Ref.\cite{LQrare} except
for $\xi^1_{RV_{1/2}}$ and $\xi^1_{RS_{1/2}}$: in these cases we have 
an analytical factor 2 of discrepancy with \cite{LQrare}.
(In particular our expressions for the bounds on 
$\xi_{RV_{1/2}}^1$ and $\xi_{RS_{1/2}}^1$ are half of
the corresponding ones in \cite{LQrare}).
However our results are consistent with the effective
Hamiltonian and the approximation\footnote{This approximation mainly
consists in assuming that the matrix elements, 
corresponding to the amplitudes $\mu~Ti \to e~Ti$ and $\mu~Ti \to 
\nu_{\mu}~Ti^{\prime}$, are comparable.} used in \cite{LQrare} 
for calculating the hadronic matrix elements.

Now we analyze the results in tables [\ref{tabS}]-[\ref{tabDV}]
for the upper bounds\footnote{Note that, in our approach, the bounds 
on $\xi^{i=2}_{LQ}$ and $\xi^{i=3}_{LQ}$ turn out to be the same.} 
on $\xi^{i=2,3}_{LQ}$ which come from the $\mu\to 3e$ decay
where the current experimental upper limit on the 
branching ratio is $B_R^{exp}(\mu\to 3e)<10^{-12}$ at $90\%$ CL \cite{PDG}.
These bounds are obtained in the approximation of 
large LQ mass limit and so by setting to zero
the $\Delta_{V,S}$ functions in (\ref{BRmu3eV})-(\ref{BRmu3eS})
which are of order $O(m_t^2/m_{LQ}^2)$. 
Moreover the following values of the quark masses are used:
$m_s=150$MeV, $m_c=1.5$GeV, $m_b=5$GeV,
and $m_t=175$GeV, these correspond to the central values 
of the allowed ranges \cite{PDG}.
From these results we see that the current upper limits on
$\mu\to 3e$ can set
bounds on $\xi_{LQ}^{2,3}$ which are at the level of 
$\xi_{LQ}^{2,3} < O(10^{-5}-10^{-4})$.
These constraints are consistent with the main approximation used in our
analysis which neglects $\xi_{LQ}^{1}$ with respect to $\xi_{LQ}^{(2,3)}$.
Note that, due to the unitarity of the $\lambda_{LQ}$ matrices, the logarithmic
dependence of the LQ mass in the right hand side of 
Eqs.(\ref{BRmu3eV})--(\ref{BRmu3eS})
disappears and it is replaced by the logarithmic dependence of the 
corresponding quark masses ratio $\log{(m_{q_2}/m_{q_3})}$
involving only the second and third generation.
Therefore, due to the logarithmic dependence upon the quark masses 
in Eqs.(\ref{BRmu3eV})--(\ref{BRmu3eS}), 
the uncertainties which affect the bounds 
on $\xi^{i=2,3}_{LQ}$ (from $\mu\to 3e$)
induced by the uncertainties on the quark masses, are small and are of the
order of $O(10\%)$.\\
In order to estimate the reliability of 
the leading logarithmic approximation when the top mass is involved,
we plot (see Figs.[\ref{fig1}]-[\ref{fig2}]) the absolute values of the
functions
$\Delta_{V,S}(x_t,Q)$ versus the LQ mass $m_{LQ}$. 
These plots have origin in $m_{LQ}=280$ GeV which roughly corresponds to
the exclusion limit obtained at HERA\cite{H1new}
for LQ masses with couplings of electromagnetic strength.
From Figs.[\ref{fig1}]-[\ref{fig2}] one can conclude 
that these corrections are smaller than $20\%$
for moderate values of the LQ masses, in particular
(scalar) $m_{S}>600$ GeV and (vectorial) $m_{V}>400$ GeV.
For the intermediate LQ mass regions,
the corrections induced by the 
Z-penguin and box diagrams, which are of order $O(m_t^2/m_{LQ}^2)$, become
relevant and they should be included in the analysis. 
In these cases one cannot 
constrain a single variable combination as the $\xi_{LQ}^i$
and this results in a complication of the analysis.
However the excluded regions in the $m_{LQ}$
and $\lambda_{LQ}^{1i}\lambda_{LQ}^{2i}$ plane could be analyzed,
for example, by means of contour plots. A complete study for this scenario
will be presented elsewhere \cite{LQnew}.
\\
The bounds on $\xi^{i=2,3}_{LQ}$, which come from
the $\mu-e$ photon--conversion, are simply obtained by 
rescaling the corresponding results of $\mu \to 3e$ in 
tables [\ref{tabS}]-[\ref{tabDV}] by a constant factor. Clearly this factor 
depends upon the experimental upper limits on the 
$\mu-e$ conversion and $\mu\to 3e$ branching ratios. 
In particular, by inserting 
(\ref{rescaling}) into (\ref{BRmu3eV}) and (\ref{BRmu3eS}),
we get the following results for the bounds
\be
(\xi^i_{LQ})^{\mu-e}_{Ti}\simeq 
\frac{1}{5.4} (\xi^i_{LQ})^{\mu\to 3e}.
\ee
However we stress that, even though the $\mu\to 3e$ process can set 
(at present) weaker constraints than the 
$(\mu-e)_{Ti}$ conversion, the
latter is model dependent
(due to the determination of the nuclear wave functions and form factors) 
while the former is not.

In tables [\ref{tabS}]-[\ref{tabDV}] for comparison we also give
the bounds obtained from the negative searches of the $\mu\to e\gamma$ decay
where the current experimental upper limit on the 
branching ratio is $Br(\mu\to e \gamma) < 1.2\times 10^{-11}$ at $90\%$ CL
\cite{muegEXP}.
Due to the higher inverse-powers in LQ-mass dependence in 
Eqs.(\ref{BRmuegV})-(\ref{BRmuegS}), it appears natural to constrain 
the variables $\tilde{\xi}_{LQ}^{i}$.
From tables [\ref{tabS}]-[\ref{tabDV}] we see that
the bounds on $\tilde{\xi}_{LQ}^i$ with $i=2,3$ 
are at the level of $\tilde{\xi}_{LQ}^i < O(10^{-3}-10^{-2})$.
In obtaining these results the terms proportional to the
$b$ quark mass with respect to the top ones have been neglected.
Moreover in the case of the scalar LQs, in order to 
simplify the analysis, we replaced the function $\rho(x)$ 
in Eq.(\ref{BRmuegS}) by its average over the range 
$600~{\rm GeV} <m_{LQ}<2000~{\rm GeV}$.
For some bounds in the $\mu\to e\gamma$ rows,
in particular $\tilde{\xi}_{LV_0}^{2,3}$ and
$\tilde{\xi}_{RV_0}^{2,3}$, in particular
we have not shown the results (this is indicated by
the symbol $--$ appearing in the tables [\ref{tabS}]-[\ref{tabDV}]).
The reason for not giving these bounds is that, at the amplitude level, 
there is an accidental cancellation of the leading term in the
$m_q^2/m_{LQ}^2$ expansion (see Eq.(\ref{BRmuegV})).
Therefore, due to a stronger GIM suppression of the next-to-leading
contribution, we expect, in these cases, weaker bounds on masses and couplings
combinations.

For fixed values of the LQ mass we can compare the bounds
on the product of couplings $\lambda_{LQ}^{2i}\lambda_{LQ}^{1i}$ 
which come from the $\mu\to e\gamma$ and $\mu\to 3e$ decay 
(or $\mu-e$ conversion).
In particular, from the results in tables [\ref{tabS}-\ref{tabDV}], one
can draw the following conclusions: 
for a LQ mass $m_{LQ}\simeq 1$ TeV the bounds on the
$\lambda_{LQ}^{2i}\lambda_{LQ}^{1i}$ combinations from 
$\mu\to 3e$ are stronger, of roughly one order of magnitude, than 
the corresponding best ones from the $\mu\to e\gamma$ decay.
On the contrary, 
if we fix the couplings $\lambda_{LQ}^{2,3}$ to be of the order
of the electromagnetic strength, the bounds on the LQ masses 
induced by the $\mu\to 3e$ are at the level of $m_{LQ} < (2-8)$TeV 
while the same constraints from $\mu\to e\gamma$ are 
$m_{LQ} < 200$GeV--$1.3$TeV.
In this respect one can argue that the bounds obtained from $\mu\to e\gamma$ 
are weaker than in $\mu\to 3e$  or $(\mu-e)_{Ti}$.
Clearly the difference between these constraints 
becomes more noticeable in line with the increase of the LQ masses.

Now we compare the results in tables [\ref{tabS}--\ref{tabDV}]
with the corresponding best bounds on $\xi^i_{LQ}$  
obtained by other processes.
From the analysis of Ref.\cite{LQrare} we learn that 
the hadron--lepton universality and 
the ratio $R=\Gamma(\pi^+\to \bar{e}\nu)/\Gamma(\pi^+\to \bar{\mu}\nu)$ can set
severe constraints on the $\lambda_{LQ}^{21}\lambda_{LQ}^{11}/m_{LQ}^2$
variables involving the first quark generation.
This is because the LQs contribute at tree-level
to the neutron $\beta$ decay,
as well as to the $\pi^+\to \bar{e}\nu$ and $\pi^+\to \bar{\mu}\nu$,
but only via box diagrams at $\mu\to e\bar{\nu}\nu$.
Then the bounds coming from the hadron--lepton universality are obtained by 
requiring that the Fermi constant in the neutron $\beta$ decay 
does not differ significantly from the muon decay measurement.
These bounds are much weaker 
than the corresponding ones set by the $\mu-e$ conversion, 
roughly between three and four order of magnitude weaker.
Analogous conclusions hold for the bounds coming from the ratio $R$.
However the $R$ measurement is particularly effective in constraining the
product of couplings with different chiralities (this special case 
is not considered in our analysis since it involves the
product of two different couplings) at the level of 
$\lambda_{Left}^{21}\lambda_{Right}^{11}  < 
10^{-6} (m_{LQ}/(100~{\rm GeV}))^2$ \cite{LQrare}.
\\
In the $K$ and $\pi$ meson sector, there are no processes,
induced by tree-level LQ exchanges, which could strongly constrain the
$\lambda_{LQ}^{22}\lambda_{LQ}^{12}$ products \cite{LQrare}.
This is not true anymore in the $\eta$ meson sector.
Indeed the $\eta \to \mu e$ decay, due to the $\bar{s} s$ quarks
content of $\eta$, gets a tree-level LQ contribution containing the 
$\lambda_{LQ}^{22}\lambda_{LQ}^{12}$ product.
However we estimated that the bounds on $\xi^{2}_{LQ}$ obtained by
means of the experimental upper limit on the $\eta \to \mu e$ 
branching ratio (which is of order $O(10^{-6})$ \cite{PDG}) 
are of order O(1) in the same unity of
tables [\ref{tabS}--\ref{tabDV}]. So they are much weaker than the 
corresponding ones in tables [\ref{tabS}--\ref{tabDV}].
Analogous conclusions hold for the bounds on $\xi^{3}_{LQ}$.
In particular we have not found any competitive process
(with respect to $\mu-e$ conversion or $\mu \to 3e$) which can set 
better or comparable constraints on the $\xi^{3}_{LQ}$ variables.

In tables [\ref{tabtauS}]-[\ref{tabtauDV}] we give the results for the
corresponding bounds in the $\tau$ sector. In this case 
two new variables $\xi_{LQ}^{e i}\equiv 
|\lambda_{LQ}^{3 i}\lambda_{LQ}^{e i}|\times
(10^2 {\rm GeV}/m_{LQ})^2$ 
and  $\xi_{LQ}^{e i}\equiv |\lambda_{LQ}^{3 i}\lambda_{LQ}^{e i}|
\times (10^2 {\rm GeV}/m_{LQ})^2$  
(and analogous generalizations of the $\tilde{\xi}_{LQ}^i$ variables,
introduced in the $\mu$ sector, like
$\tilde{\xi}^{e i}_{LQ}$ and $\tilde{\xi}^{\mu i}_{LQ}$) are defined.
The $\xi_{LQ}^{e 1}$ and $\xi_{LQ}^{e 2}$ are more strongly constrained by the
processes $\tau\to \pi^0 e$ and $\tau\to \pi^0 \mu$ \cite{PDG} whose 
effective Hamiltonians are induced at tree-level by the LQs.
In particular for a four-fermion vertex of the form
\be
\frac{\lambda_{LQ}^{3 1} \lambda_{LQ}^{l 1}}{m_{LQ}^2}
\left(\bar{l}\gamma^{\mu}P^l\tau\right)
\left(\bar{q}_1\gamma^{\mu}P^q q_1\right)
\ee
one obtains\footnote{
In obtaining the Eq.(\ref{tautree}) the following approximations for the matrix
elements have been used \cite{LQrare}
$\langle 0|\bar{u}\gamma^{\mu}\gamma_5 u|\pi^0\rangle\simeq
 \langle 0|\bar{d}\gamma^{\mu}\gamma_5 d|\pi^0\rangle\simeq
 \langle 0|\bar{d}\gamma^{\mu}\gamma_5 u|\pi^+\rangle$.}\cite{LQrare}
\be
\xi_{LQ}^{l 1} < 2\sqrt{2}(G_F\times (10^2 {\rm GeV})^2)\cos{\theta_c}
\sqrt{\frac{B_R^{exp}(\tau^-\to \pi^0 l^-)}{B_R^{exp}(\tau^-\to \pi^- \nu)}},
\label{tautree}
\ee
where $q_1=U,D$ quarks, $\theta_c$ is the Cabibbo angle, $l=e,\mu$, 
and the central values of the experimental
branching ratio is $B_R^{exp}(\tau\to \pi^-\nu)=11.08\%$ \cite{PDG}.
The bounds in the $i=2,3$ sectors can be simply obtained by rescaling the 
corresponding ones in the $\mu$ sector 
in tables [\ref{tabtauS}]-[\ref{tabtauDV}] by means
of the Eqs.(\ref{BRtaulg}),(\ref{BRtau3l}).
In obtaining these bounds the following experimental upper limits at $90\%$ CL
have been used \cite{PDG}:
$B_R^{exp}(\tau\to \pi^0 e)<3.7 \times 10^{-6}$, 
$B_R^{exp}(\tau\to \pi^0 \mu)<4\times 10^{-6}$,
$B_R^{exp}(\tau\to 3e)<2.9\times 10^{-6}$,
$B_R^{exp}(\tau\to 3\mu)<1.9\times 10^{-6}$, 
$B_R^{exp}(\tau\to e \gamma)<2.7\times 10^{-6}$, and
$B_R^{exp}(\tau\to \mu \gamma)<3.0\times 10^{-6}$.
Due to a much lower experimental sensitivity in the $\tau$ branching ratios,
we see that the bounds in tables [\ref{tabtauS}]-[\ref{tabtauDV}] are 
between three and four order of magnitude weaker than  
the corresponding ones in the $\mu$ sector.

We discuss now the improvements of the LQ 
bounds in the muon sector which could be reached 
at the present and future muon experiments.
In this respect it is convenient to introduce the {\it improvement} 
factor $B=\sqrt{{\rm BR}^{exp}_{curr}/{\rm BR}^{exp}_{fut}}$
where ${\rm BR}^{exp}_{curr}$ and ${\rm BR}^{exp}_{fut}$ 
are the current and future upper limits on the branching ratios, respectively.
In the $\mu \to e \gamma$ sector the new bounds are obtained 
by dividing the current ones by $\sqrt{B}$  while in the 
$\mu \to 3e$ or $\mu-e$ conversion they should be divided by $B$.
In the sector of the experimental searches for $\mu^{+} \to e^{+}\gamma$ it
seems feasible, by using polarized muons (which are useful for suppressing the
backgrounds \cite{okada}), to reach the
sensitivity of about $10^{-14}$ on the branching ratio \cite{newpsi}. 
This is converted into an improvement factor $\sqrt{B}\simeq 8$
for the bounds from $\mu \to e \gamma$.
The final analysis of the current SINDRUM II experiment at Paul Scherrer
Institute (PSI) on the $\mu-e$ conversion \cite{scharf}
will reach a sensitivity of $10^{-14}$ on the branching ratio. This will
give an improvement factor $B\simeq 8$ in the new bounds
from the $\mu-e$ conversion.

A recent proposal for the  
$\mu-e$ conversion experiment (MECO) \cite{meco}
at Brookhaven National Laboratory (BNL)
will permit a sensitivity on the branching ratio better than $10^{-16}$. 
This sensitivity is translated into an improvement factor $B\simeq 80$ 
for the LQ bounds from the $\mu-e$ conversion.
In the context of the $\mu \to 3e$ decay (apparently) 
there is not any proposal, at the present and future 
muon facilities machines, for improving the sensitivity on
this branching ratio.
\section{Conclusions}
In this article we perform a model independent analysis so as to constrain
the LQs models ($B$ and $L$ conserving) 
in the sector of rare FC leptonic processes; this is done
by means of the $\mu\to e\gamma$, $\mu \to 3e$ decays (and analogous decays 
in the $\tau$ sector) and the $\mu-e$ conversion in nuclei. 
In our analysis we assume that the LQ couplings
$\lambda_{LQ}^{l_iq_j}$ (where $l_i$ and $q_j$ indicate the generation numbers
of lepton and quarks respectively) are unitary and real matrices.
In order to set bounds on the LQ couplings and masses 
we find it convenient to introduce the following variables:
$\xi^{i}_{LQ}=|\lambda_{LQ}^{2i} \lambda_{LQ}^{1i}|\times 
(10^2 {\rm GeV}/m_{LQ})^2$ in the $\mu$ sector,
$\xi^{e i}_{LQ}=|\lambda_{LQ}^{3i} \lambda_{LQ}^{1i}|\times 
(10^2 {\rm GeV}/m_{LQ})^2$ and
$\xi^{\mu i}_{LQ}=|\lambda_{LQ}^{3i} \lambda_{LQ}^{2i}|\times 
(10^2 {\rm GeV}/m_{LQ})^2$ in the $\tau$ one.

The $\mu-e$ conversion in nuclei, as shown in \cite{LQrare},
is the best process for constraining the FC leptoquark couplings involving
the first generation of quarks, namely the $\xi^{1}_{LQ}$, this is because
the relevant effective Hamiltonian is induced at tree-level.
The couplings involving the second and third quark generations can also
be constrained by means of the one-loop contributions to the
$\mu\to e\gamma$, $\mu \to 3e$ decays, and the 
$\mu-e$ conversion in nuclei.
We show that the best of these processes where 
to strongly constrain the variables $\xi^{2,3}_{LQ}$
are the $\mu \to 3e$ and the $(\mu-e)_{Ti}$.
On the contrary the $\mu\to e\gamma$ decay can set much weaker constraints.
This is because the former
receive large logarithms ($\log{(m_q/m_{LQ})}$)
enhancements at the amplitude level while  
the latter does not.
In particular the $\mu \to 3e$ decay and $\mu-e$ conversion in nuclei
(mediated by the photon-conversion mechanism)
get the leading contributions from the so called 
photon-penguin diagrams which, in the large LQ mass limit,
are proportional to $\lambda_{LQ}^2/m_{LQ}^2 \times \log{(m_q/m_{LQ})}$.
On the other hand the amplitude of the $\mu \to e\gamma$ is 
proportional (in the large LQ mass limit) to $\lambda_{LQ}^2 m_q^2/m_{LQ}^4$.
The same considerations regarding the log enhancements 
hold for the corresponding processes in the $\tau$ sector.

The log enhancements in the photon-penguins 
are known in the literature \cite{log1}-\cite{log2}, 
but they have not been applied in this context.
In particular in the analysis of Ref.\cite{LQrare} 
these diagrams were not taken into account. 
As a consequence, in Ref.\cite{LQrare}, 
the $\mu\to 3e$ branching ratio has been by far underestimated.
Moreover in Ref.\cite{LQrare} the scalar LQ couplings were not assumed to be 
unitary, thus allowing for large scalar LQ contributions 
to the $\mu \to e\gamma$ branching ratio.
\\
The complete list of the bounds which we establish can be found 
in tables [\ref{tabS}]-[\ref{tabtauDV}] in both the $\mu$ and $\tau$ sectors.
We next briefly outline the general trend of our results
and describe the impact of 
the future experimental sensitivities on the muon branching ratios on our
bounds.
\begin{itemize}
\item
The best constraints on all the $\xi^{i}_{LQ}$ variables
that we establish come from the current experimental 
upper limit on the $(\mu-e)_{Ti}$ branching ratio \cite{mueEXP}.
In particular, in both scalar and vectorial LQ sectors, 
the strongest bounds on $\xi^{1}_{LQ}$ are at the level of
$\xi^{1}_{LQ} < 10^{-7}$ while
the strongest  bounds on $\xi^{2,3}_{LQ}$ are weaker and are 
at the level of $\xi^{2,3}_{LQ} < 10^{-5}$.
The current experimental upper limits on the $\mu\to 3e$ decay can also 
set strong constraints on the $\xi_{LQ}^{2,3}$
variables, however they are roughly a factor 5 weaker with 
respect to the corresponding ones in $\mu-e$ sector. 
\item
The bounds obtained 
from the $\mu\to 3e$ decay can be calculated with high accuracy 
in perturbation theory, whereas the corresponding ones 
from the $\mu-e$ conversion in nuclei suffer from the problem
of model dependent calculations in the nuclear sector.
\item
Because of a lower sensitivity in the $\tau$ experimental branching ratios
\cite{PDG}, 
the bounds obtained from the rare $\tau$ decays are roughly 
four order of magnitude weaker than the bounds in the $\mu$ sector.
In particular the best bounds on $\xi^{e 1}_{LQ}$ and 
$\xi^{\mu 1}_{LQ}$ are at the level of
$\xi^{\mu 1}_{LQ} < 10^{-3}$ and are set by
the decays $\tau\to \pi^0 e$ and $\tau\to \pi^0 \mu$, respectively.
The best 
bounds on $\xi^{e (2,3)}_{LQ}$ and $\xi^{\mu (2,3)}_{LQ}$ come from the
$\tau\to 3e$ and $\tau\to 3\mu$ decays respectively,
and are at the level of $O(10^{-1})$.
\item
Ultimately the current experiment on the  $(\mu-e)_{Ti}$ conversion at PSI 
\cite{mueEXP}-\cite{newpsi} will reach a sensitivity on
the branching ratio that will enable us to improve the bounds in tables
[\ref{tabS}]-[\ref{tabDV}] by a factor $\simeq 8$.
However a new proposal (called MECO) for the $(\mu-e)_{Ti}$ conversion 
at BNL \cite{meco},
could reach a sensitivity on the branching ratio that will considerably
improve the bounds
in tables [\ref{tabS}]-[\ref{tabDV}]  by a factor $\simeq 80$.
\end{itemize}
\section*{Acknowledgments}
We gratefully acknowledge discussions with 
A. De Rujula, C. Di Cairano, 
B. Gavela, G. Isidori, 
C. Munoz, L. Silvestrini, A. Van der Schaaf, and D. Zanello.
I acknowledge the financial supports of the TMR network, project
``Physics beyond the standard model'', FMRX-CT96-0090 and the partial
financial support of the CICYT project ref. AEN97-1678.
\section*{Appendix}
We here give the analytical expressions for the
functions $\Delta_{S,V}(x_t,Q)$
which appear in the branching ratios in 
Eqs.(\ref{BRmuegV})-(\ref{BRmuegS}), where
$x_t=m_t^2/m^2_{LQ}$ and $Q$ can be $Q=R_{LQ}$ or $Q=-R_{LQ}$, with
$R_{LQ}\equiv Q_{LQ}/Q_{U}$.
In Eqs.(\ref{BRmuegV})-(\ref{BRmuegS}) 
the dependence by $Q$ take only two
values, namely $Q=(1/2, -5/2)$. These functions, which are
of order $O(x)$, take into account the exact dependence on 
the top mass in the photon--penguin diagrams and give the
percentage difference between the leading logarithm approximation 
(which consists in neglecting 
the terms of order $O(x_t)$   with respect to $\log(x_t)$)
and the exact result.
Their expressions are given by
\bea
\Delta_V(x,Q)&=&\frac{x\left(-18+11 x+x^2+Q (-12-x+7 x^2)\right)}{8(1-x)^3
\log{(x)}}\no
&-&\frac{x^2\left(15-16x+4x^2+Q(12-10x+x^2)\right)}{4(x-1)^4}\no
\Delta_S(x,Q)&=&\frac{x\left(-19+41x-16x^2+Q(-1+5x+2x^2)\right)}
{12(x-1)^3\log{(x)}}\no
&-&\frac{x\left(-5+12x+(Q-8)x^2+2x^3\right)}{2(x-1)^4}.
\eea
In order to estimate the reliability of the leading log approximation,  
in figures [\ref{fig1}]-[\ref{fig2}] we plot
the absolute values of above functions versus the leptoquark mass
in both the scalar and vectorial cases.
\begin{table}
\begin{center}
\begin{tabular}{|r||c|c|}
\hline 
${\rm LQ}$
& $Q_{LQ}$ 
\\ \hline 
   \hline $S_0 $ & $ -1/3 $ 
\\
$\tilde{S}_0 $ & $ -4/3 $ 
\\ \hline 
   \hline $(S_{1/2}^U~,~S_{1/2}^D)$ & $ (-2/3,-5/3)$ 
\\
   \hline $(\tilde{S}_{1/2}^U~,~\tilde{S}_{1/2}^D)$ & $ (1/3,-2/3)$ 
\\ \hline 
   \hline $(S_1^1~,~S_1^2~,~S_1^3)$ & $ (-4/3,2/3,-1/3) $ 
\\ \hline 
   \hline $V_0 $ & $ -2/3 $ 
\\
$\tilde{V}_0 $ & $ -5/3 $ 
\\ \hline 
   \hline $(V_{1/2}^U~,~V_{1/2}^D)$ & $ (-1/3,-4/3)$ 
\\
   \hline $(\tilde{V}_{1/2}^U~,~\tilde{V}_{1/2}^D)$ & $ (2/3,-1/3)$ 
\\ \hline 
   \hline $(V_1^1~,~V_1^2~,~V_1^3)$ & $ (1/3,-2/3,-5/3) $ 
\\ \hline
\end{tabular}
\caption[]{\small Electromagnetic charges of the scalar (S) and vectorial (V)
LQs in unit of $e$.}
\label{LQcharge}
\end{center}
\end{table}
\begin{table}
\begin{center}
\begin{tabular}{|r|c||c|c|c|c|}
\hline 
${\rm {\bf Vector}}$
& $i$
& $(\xi_{LV_0}^{i})^B$ 
& $(\xi_{RV_0}^{i})^B$ 
& $(\xi_{R\tilde{V}_0}^{i})^B$ 
\\   \hline $(\mu-e)_{Ti}$ & 
1 & $2.6\times 10^{-7}$ & $2.6\times 10^{-7}$ & $2.6\times 10^{-7}$
\\   \hline $(\mu-e)_{Ti}$ &
2,3 & $1.5\times 10^{-5}$ & $1.5\times 10^{-5}$ & 
    $4.6\times 10^{-6}$
\\   \hline $\mu\to 3e$ &
2,3 & $8.0\times 10^{-5}$ & $8.0\times 10^{-5}$ & 
    $2.5\times 10^{-5}$
\\   \hline $\mu\to e\gamma$ &
2,3 & $--$ & $--$ & 
    $1.7\times 10^{-3}$
\\ \hline\hline 
${\rm {\bf Scalar}}$
& $i$
& $(\xi_{LS_0}^{i})^B$ 
& $(\xi_{RS_0}^{i})^B$ 
& $(\xi_{R\tilde{S}_0}^{i})^B$ 
\\   \hline $(\mu-e)_{Ti}$ & 
1 & $5.2\times 10^{-7}$ 
& $5.2\times 10^{-7}$ & $5.2\times 10^{-7}$ 
\\   \hline $(\mu-e)_{Ti}$ &
2,3 & $9.2\times 10^{-6}$ & 
$9.2\times 10^{-6}$ & $3.0\times 10^{-5}$
\\   \hline $\mu\to 3e$ &
2,3 & $5.0\times 10^{-5}$ & 
$5.0\times 10^{-5}$ & $1.6\times 10^{-4}$
\\   \hline $\mu\to e\gamma$ &
2,3 & $2.3\times 10^{-3}$ & $2.3\times 10^{-3}$ & 
    $4.5\times 10^{-2}$
\\ \hline
\end{tabular}
\caption[]{\small Numerical upper bounds for the  
vectorial and  scalar $SU(2)_L$--{\it singlet}--LQ variables
$\xi_{LQ}^{i}\equiv |\lambda_{LQ}^{2 i}\lambda_{LQ}^{1 i}|\times
\left(10^2{\rm GeV}/m_{LQ}\right)^2 < (\xi^i_{LQ})^B$
obtained from the experimental upper limits on the $(\mu-e)_{Ti}$ 
conversion rate and $\mu\to 3e$ decay.
In the $\mu\to e\gamma$ rows the variables which are constrained
are $\tilde{\xi}_{LQ}^{i}\equiv 
\sqrt{|\lambda_{LQ}^{2 i}\lambda_{LQ}^{1 i}|}\times
\left(10^2{\rm GeV}/m_{LQ}\right)^2 < (\xi^i_{LQ})^B$.
Note that the symbol $--$ stands for weaker constraints, see the text.
Everywhere $(\xi^2_{LQ})^B=(\xi^3_{LQ})^B$.}
\label{tabS}
\end{center}
\end{table}
\begin{table}
\begin{center}
\begin{tabular}{|r|c||c|c|c|c|c|}
\hline 
${\rm {\bf Vector}}$
& $i$
& $(\xi_{LV_{1/2}}^{i})^B$ 
& $(\xi_{RV_{1/2}}^{i})^B$ 
& $(\xi_{L\tilde{V}_{1/2}}^{i})^B$ 
& $(\xi_{LV_1}^{i})^B$ 
\\   \hline $(\mu-e)_{Ti}$ & 
1 & $2.6\times 10^{-7}$ 
& $1.3\times 10^{-7}$ & $2.6\times 10^{-7}$ & $8.5\times 10^{-8}$
\\   \hline $(\mu-e)_{Ti}$ &
2,3 & $1.5\times 10^{-5}$ & 
    $6.7\times 10^{-6}$ & $4.6\times 10^{-6}$ & 
    $2.7\times 10^{-6}$
\\   \hline $\mu\to 3e$ &
2,3 & $8.0\times 10^{-5}$ & 
    $3.7\times 10^{-5}$ & $2.5\times 10^{-5}$ & 
    $1.5\times 10^{-5}$
\\   \hline $\mu\to e\gamma$ &
2,3 & $8.4\times 10^{-2}$ & 
    $2.9\times 10^{-3}$ & $2.9\times 10^{-3}$ & 
    $1.2\times 10^{-3}$
\\ \hline\hline
${\rm {\bf Scalar}}$
& $i$
& $(\xi_{LS_{1/2}}^{i})^B$ 
& $(\xi_{RS_{1/2}}^{i})^B$ 
& $(\xi_{L\tilde{S}_{1/2}}^{i})^B$ 
& $(\xi_{LS_1}^{i})^B$ 
\\   \hline $(\mu-e)_{Ti}$ & 
1 & $5.2\times 10^{-7}$ & $2.6 \times 10^{-7}$ & $5.2 \times 10^{-7}$ 
& $1.7 \times 10^{-7}$ 
\\   \hline $(\mu-e)_{Ti}$ &
2,3 & $9.2\times 10^{-6}$ & $1.3\times 10^{-5}$ & $3.0\times 10^{-5}$ & 
    $2.5\times 10^{-5}$
\\   \hline $\mu\to 3e$ &
2,3 & $5.0\times 10^{-5}$ & $7.3\times 10^{-5}$ & $1.6\times 10^{-4}$ & 
    $1.3\times 10^{-4}$
\\   \hline $\mu\to e\gamma$ &
2,3 & $1.7\times 10^{-3}$ & $1.7\times 10^{-3}$ & $5.1\times 10^{-2}$ & 
    $2.3\times 10^{-3}$
\\ \hline
\end{tabular}
\caption[]{\small Numerical bounds as in Table [\ref{tabS}] which here are
relative to the $SU(2)_L$--{\it doublet--} and $SU(2)_L$--{\it vector}--LQs.}
\label{tabDV}
\end{center}
\end{table}
\begin{table}
\begin{center}
\begin{tabular}{|r|c||c|c|c|c|}
\hline 
${\rm {\bf Vector}}$
& $i$
& $(\xi_{LV_0}^{e i})^B$ 
& $(\xi_{RV_0}^{e i})^B$ 
& $(\xi_{R\tilde{V}_0}^{e i})^B$ 
\\   \hline $\tau \to \pi^0 e$ & 
1 & $1.9\times 10^{-3}$ & $1.9\times 10^{-3}$ & $1.9\times 10^{-3}$
\\   \hline $\tau\to 3e$ &
2,3 & $3.3\times 10^{-1}$ & $3.3\times 10^{-1}$ & 
    $1.0\times 10^{-1}$
\\   \hline $\tau\to e\gamma$ &
2,3 & $--$ & $--$ & 
    $5.7\times 10^{-2}$
\\ \hline\hline
${\rm {\bf Vector}}$
& $i$
& $(\xi_{LV_0}^{\mu i})^B$ 
& $(\xi_{RV_0}^{\mu i})^B$ 
& $(\xi_{R\tilde{V}_0}^{\mu i})^B$ 
\\   \hline $\tau \to \pi^0 \mu$ &
1 & $1.9\times 10^{-3}$ & $1.9\times 10^{-3}$ & $1.9\times 10^{-3}$
\\   \hline $\tau\to 3\mu$ &
2,3 & $2.7\times 10^{-1}$ & $2.7\times 10^{-1}$ & 
    $8.3\times 10^{-2}$
\\   \hline $\tau\to \mu\gamma$ &
2,3 & $--$ & $--$ & $5.8\times 10^{-2}$
\\ \hline\hline 
${\rm {\bf Scalar}}$
& $i$
& $(\xi_{LS_0}^{e i})^B$ 
& $(\xi_{RS_0}^{e i})^B$ 
& $(\xi_{R\tilde{S}_0}^{e i})^B$ 
\\   \hline $\tau \to \pi^0 e$ & 
1 & $3.7\times 10^{-3}$ & $3.7\times 10^{-3}$ & $3.7\times 10^{-3}$
\\   \hline $\tau\to 3e$ &
2,3 & $2.0\times 10^{-1}$ & $2.0\times 10^{-1}$ & 
$6.6\times 10^{-1}$
\\   \hline $\tau\to e\gamma$ &
2,3 & $7.7\times 10^{-2}$ & $7.7\times 10^{-2}$ & $1.5$
\\ \hline\hline
${\rm {\bf Scalar}}$
& $i$
& $(\xi_{LS_0}^{\mu i})^B$ 
& $(\xi_{RS_0}^{\mu i})^B$ 
& $(\xi_{R\tilde{S}_0}^{\mu i})^B$ 
\\   \hline $\tau \to \pi^0 \mu$ &
1 & $3.9\times 10^{-3}$ & $3.9\times 10^{-3}$ & $3.9\times 10^{-3}$
\\   \hline $\tau\to 3\mu$ &
2,3 & $1.7\times 10^{-1}$ & $1.7\times 10^{-1}$ & 
    $5.3\times 10^{-1}$
\\   \hline $\tau\to \mu\gamma$ &
2,3 & $7.9\times 10^{-2}$ & $7.9\times 10^{-2}$ & 
    $1.6$
\\ \hline
\end{tabular}
\caption[]{\small Numerical upper bounds for the 
vectorial and scalar $SU(2)_L$--{\it singlet}--LQ variables
$\xi_{LQ}^{e i}\equiv |\lambda_{LQ}^{3 i}\lambda_{LQ}^{1 i}|\times
\left(10^2{\rm GeV}/m_{LQ}\right)^2 < (\xi^{e i}_{LQ})^B$
and 
$\xi_{LQ}^{\mu i}\equiv |\lambda_{LQ}^{3 i}\lambda_{LQ}^{2 i}|\times
\left(10^2{\rm GeV}/m_{LQ}\right)^2 < (\xi^{\mu i}_{LQ})^B$
obtained from the experimental upper limits on the $\tau \to \pi^0 e$,
$\tau \to 3e$ and $\tau \to \pi^0 \mu$, $\tau \to 3\mu$, respectively.
In the 
$\tau\to e\gamma$ and  $\tau\to \mu\gamma$ rows, the variables which 
are constrained are $\tilde{\xi}_{LQ}^{e i}\equiv 
\sqrt{|\lambda_{LQ}^{3 i}\lambda_{LQ}^{1 i}|}\times
\left(10^2{\rm GeV}/m_{LQ}\right)^2 < (\xi^{e i}_{LQ})^B$
and $\tilde{\xi}_{LQ}^{\mu i}\equiv 
\sqrt{|\lambda_{LQ}^{3 i}\lambda_{LQ}^{2 i}|}\times
\left(10^2{\rm GeV}/m_{LQ}\right)^2 < (\xi^{e i}_{LQ})^B$, respectively.
Note that the symbol $--$ stands for weaker constraints, see the text.
Everywhere $(\xi^{e 2}_{LQ})^B=(\xi^{e 3}_{LQ})^B$ and 
$(\xi^{\mu 2}_{LQ})^B=(\xi^{\mu 3}_{LQ})^B$}
\label{tabtauS}
\end{center}
\end{table}

\begin{table}
\begin{center}
\begin{tabular}{|r|c||c|c|c|c|c|}
\hline 
${\rm {\bf Vector}}$
& $i$
& $(\xi_{LV_{1/2}}^{e i})^B$ 
& $(\xi_{RV_{1/2}}^{e i})^B$ 
& $(\xi_{L\tilde{V}_{1/2}}^{e i})^B$ 
& $(\xi_{LV_1}^{e i})^B$ 
\\   \hline $\tau \to \pi^0 e$ &
1   & $1.9\times 10^{-3}$ & $9.3\times 10^{-4}$ & 
      $1.9\times 10^{-3}$ & $6.2\times 10^{-4}$
\\   \hline $\tau\to 3e$ &
2,3 & $3.3\times 10^{-1}$ & $1.5\times 10^{-1}$ & 
      $1.0\times 10^{-1}$ & $6.1\times 10^{-2}$
\\   \hline $\tau\to e\gamma$ &
2,3 & $2.8$ & $9.8\times 10^{-2}$ & 
      $9.8\times 10^{-2}$ & $4.0\times 10^{-2}$
\\ \hline\hline 
${\rm {\bf Vector}}$
& $i$
& $(\xi_{LV_{1/2}}^{\mu i})^B$ 
& $(\xi_{RV_{1/2}}^{\mu i})^B$ 
& $(\xi_{L\tilde{V}_{1/2}}^{\mu i})^B$ 
& $(\xi_{LV_1}^{\mu i})^B$ 
\\   \hline $\tau \to \pi^0 \mu$ &
1   & $1.9\times 10^{-3}$ & $9.7\times 10^{-4}$ & 
      $1.9\times 10^{-3}$ & $6.4\times 10^{-4}$
\\   \hline $\tau\to 3\mu$ &
2,3 & $2.7\times 10^{-1}$ & $1.2\times 10^{-1}$ & 
      $8.3\times 10^{-2}$ & $4.9\times 10^{-2}$
\\   \hline $\tau\to \mu\gamma$ &
2,3 & $2.9$ & $1.0\times 10^{-1}$ & 
      $1.0\times 10^{-1}$ & $4.1\times 10^{-2}$
\\ \hline\hline 
${\rm {\bf Scalar}}$
& $i$
& $(\xi_{LS_{1/2}}^{e i})^B$ 
& $(\xi_{RS_{1/2}}^{e i})^B$ 
& $(\xi_{L\tilde{S}_{1/2}}^{e i})^B$ 
& $(\xi_{LS_1}^{e i})^B$ 
\\   \hline $\tau \to \pi^0 e$ &
1   & $3.7\times 10^{-3}$ & $1.9\times 10^{-3}$ & 
      $3.7\times 10^{-3}$ & $1.2\times 10^{-3}$
\\   \hline $\tau\to 3e$ &
2,3 & $2.0\times 10^{-1}$ & $3.0\times 10^{-1}$ & 
      $6.6\times 10^{-1}$ & $5.5\times 10^{-1}$
\\   \hline $\tau\to e\gamma$ &
2,3 & $5.7\times 10^{-2}$ & $5.7\times 10^{-2}$ & 
      $1.7$ & $7.7\times 10^{-2}$
\\ \hline\hline 
${\rm {\bf Scalar}}$
& $i$
& $(\xi_{LS_{1/2}}^{\mu i})^B$ 
& $(\xi_{RS_{1/2}}^{\mu i})^B$ 
& $(\xi_{L\tilde{S}_{1/2}}^{\mu i})^B$ 
& $(\xi_{LS_1}^{\mu i})^B$ 
\\   \hline $\tau \to \pi^0 \mu$ &
1   & $3.9\times 10^{-3}$ & $1.9\times 10^{-3}$ & 
      $3.9\times 10^{-3}$ & $1.3\times 10^{-3}$
\\   \hline $\tau\to 3\mu$ &
2,3 & $1.7\times 10^{-1}$ & $2.4\times 10^{-1}$ & 
      $5.3\times 10^{-1}$ & $4.4\times 10^{-1}$
\\   \hline $\tau\to \mu\gamma$ &
2,3 & $5.8\times 10^{-2}$ & $5.8\times 10^{-2}$ & 
      $1.8$ & $7.9\times 10^{-2}$
\\ \hline
\end{tabular}
\caption[]{\small Numerical bounds as in table [\ref{tabtauS}]
which here are relative to 
the $SU(2)_L$--{\it doublet--} and $SU(2)_L$--{\it vector}--LQs.}
\label{tabtauDV}
\end{center}
\end{table}
\begin{figure}[b]
\centerline{\psfig{figure=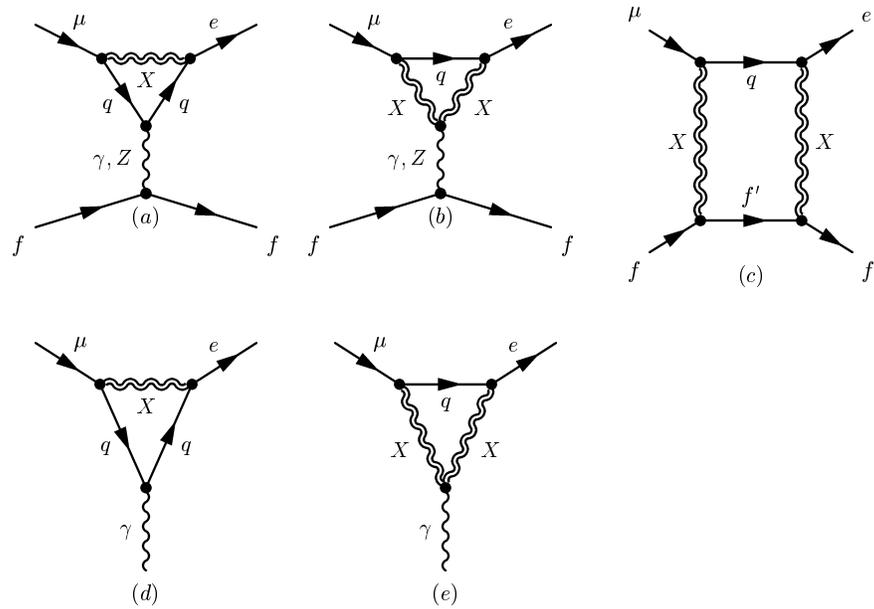,height=25cm,angle=0}}
\vspace{-12cm}
\caption{{\small 
The photon ($\gamma$)-- and Z--penguin (a)--(b) and the box (c) 
diagrams for the $\mu\to e f \bar{f}$ process,
and the magnetic-penguin (d)--(e) diagrams 
for the $\mu\to e \gamma$ process, in the LQ model,
where $f$, $q$, and $X$
indicate a general external fermion, quarks and LQs respectively.
In addition to (a)--(b) there are also the diagrams 
with the self-energy insertions (not included in this figure),
where the $\gamma$ or Z is attached on the external $e$ and $\mu$, 
which contribute to the $\mu\to e f \bar{f}$ amplitude.
}}
\label{diag}  
\end{figure}
\begin{figure}[t]
\centerline{\epsfig{figure=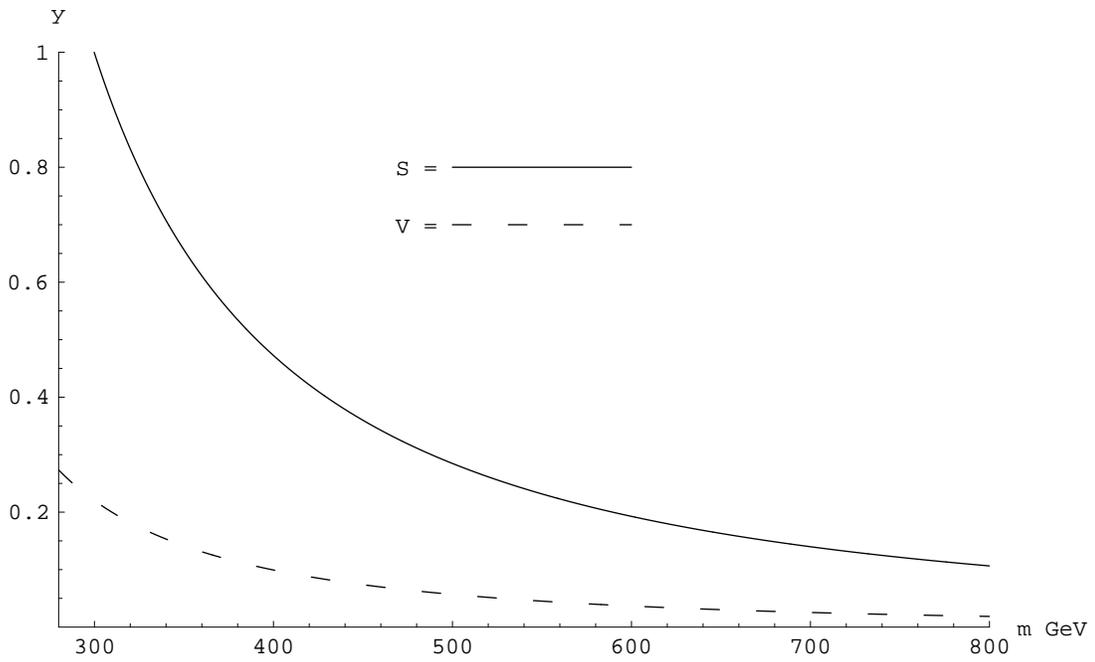,height=20cm,angle=0}}
\vspace{-5cm}
\caption{{\small 
Plots for $y=|\Delta_S(x_t,Q)|$ (continuous) 
and $y=|\Delta_V(x_t,Q)|$ (dashed) functions, with $x_t=m_t^2/m_{LQ}^2$
and $Q=-5/2$, versus $m=m_{LQ}$ in GeV.}}
\label{fig1}  
\end{figure}
\begin{figure}[t]
\centerline{\epsfig{figure=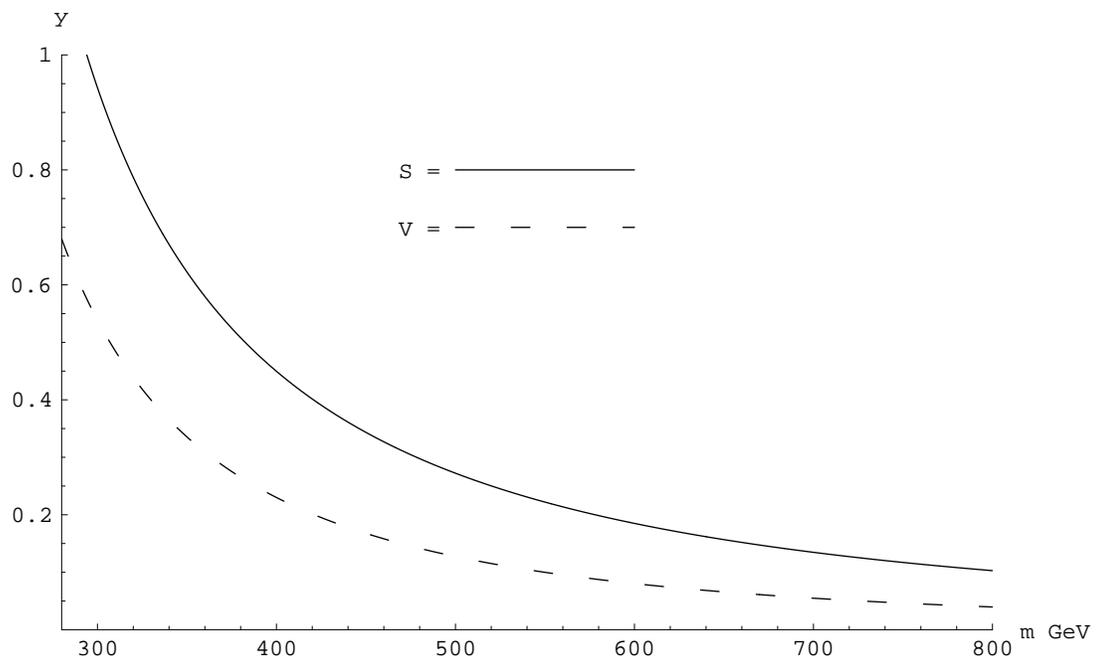,height=20cm,angle=0}}
\vspace{-5cm}
\caption{{\small 
The same plots as in Fig.[\ref{fig1}] but for $Q=1/2$.}}
\label{fig2}  
\end{figure}
\newpage

\end{document}